\documentclass[reprint,twocolumn,prx,superscriptaddress]{revtex4-2}
\usepackage[utf8]{inputenc}
\usepackage[T1]{fontenc}
\usepackage{amsmath}
\usepackage{amsfonts}
\usepackage{physics}
\usepackage[caption=false]{subfig}
\usepackage{bm}
\usepackage{amssymb}
\usepackage[mathscr]{eucal}
\usepackage{graphicx}
\usepackage{xcolor}
\usepackage[unicode=true,pdfusetitle,
 bookmarks=true,bookmarksnumbered=false,bookmarksopen=false,
 breaklinks=false,pdfborder={0 0 1},backref=false,colorlinks=false]
 {hyperref}
\hypersetup{
 colorlinks=true,allcolors=blue}
\usepackage{xr}

\begin{document}
\title{Thermodynamic coupling of reactions via few-molecule vibrational polaritons}
\author{Arghadip Koner}
\affiliation{Department of Chemistry and Biochemistry, University of California
San Diego, La Jolla, California 92093, USA}
\author{Matthew Du}
\affiliation{Department of Chemistry, University of Chicago, 5735 S Ellis Ave, Chicago, Illinois 60637, USA}
\author{Sindhana Pannir-Sivajothi}
\affiliation{Department of Chemistry and Biochemistry, University of California
San Diego, La Jolla, California 92093, USA}
\author{Randall H. Goldsmith}
\affiliation{Department of Chemistry, University of Wisconsin-Madison, Madison, Wisconsin 53706-1322, USA}
\author{Joel Yuen-Zhou}
\email{joelyuen@ucsd.edu}
\affiliation{Department of Chemistry and Biochemistry, University of California
San Diego, La Jolla, California 92093, USA}

\begin{abstract}
Interaction between light and molecular vibrations leads to hybrid light-matter states called vibrational polaritons. Even though many intriguing phenomena have been predicted for single-molecule vibrational strong coupling (VSC), several studies suggest that these effects tend to be diminished in the many-molecule regime due to the presence of dark states. Achieving single or few-molecule vibrational polaritons has been constrained by the need for fabricating extremely small mode volume infrared cavities. In this work, we propose an alternative strategy to achieve
single-molecule VSC in a cavity-enhanced Raman spectroscopy (CERS) setup, based on the physics of cavity optomechanics. We then present a scheme harnessing few-molecule VSC to thermodynamically couple two reactions, such that a spontaneous electron transfer can now fuel a thermodynamically uphill reaction that was non-spontaneous outside the cavity.\\

\end{abstract}

\maketitle


\section*{Introduction}

Strong coupling (SC) ensues when the rate of coherent energy exchange between matter degrees of freedom (DOF) and a confined electromagnetic field exceeds the losses from either of them~\cite{microcavitiesbook, Tormarev_2015, Haranrev2019,masiello_2022}. This interplay leads to the emergence of hybrid light-matter states called polaritons~\cite{raphaelrev2018, mattnoneq2021,cina2022}, which inherit properties from both the photonic and the matter constituents. For molecular systems, due to the small magnitude of the transition dipole moment of most individual molecules, SC is typically achieved by having an ensemble of $N\gg 1$ molecules interact with a cavity mode~\cite{raphaelrev2018, Hiroshirev2020}.  In this collective case, in addition to two polariton states, SC leads to ($N-1$) dark states which are predominantly molecular in character~\cite{mattdark2022}. In both the electronic and vibrational regimes, harnessing these hybrid light-matter states has led to the emergence of a plethora of polariton-based devices~\cite{Fraser_2017} such as amplifiers~\cite{beriniamp2012, Jamadiamp2018}, tunneling diodes~\cite{Blochdiode2013}, routers~\cite{Felix_rou_2015}, and ultrafast switches~\cite{Amo2010, Chen_sw_2022, vahid_2021}; and novel phenomena like  enhanced energy and charge transport~\cite{Wang2021,mattparet_2018,groenhoff_2018},  modification and control of a chemical reaction without external pumping ~\cite{Thomas2019, Hirai2020}, and remote catalysis~\cite{matt_remote_2019}.\\

Theoretical models of polaritons often use a single molecule with a collective superradiant coupling to the cavity to explain the experimentally observed effects of collective SC on physical and chemical phenomena~\cite{Rubio_2020, Rubio1_2021, Rubio2_2021, pelton_2020}. However, several theoretical studies, that account for the large number of molecules coupled to the cavity, suggest that SC could be rendered less effective in the collective regime owing to the entropic penalty from the dark states~\cite{mattnoneq2021,taoli_2022}. For enhanced polaritonic effects, the state-of-the-art is either to use polariton condensates~\cite{keeling_rev_2020, deliberato_2017} or to achieve single-molecule SC\cite{Chikkaraddy2016}. In the electronic regime, both polariton condensation~\cite{laguodakis2019, Kena-Cohen2010, Baumberg2008, Ghosh2020} and single-molecule SC~\cite{Chikkaraddy2016, pelton_2019} have been achieved.
There have been theoretical proposals of ways to achieve a vibrational polariton condensate ~\cite{sindhana2022}. However, to the best of our knowledge, in the vibrational regime, neither condensation nor single-molecule SC has yet been experimentally demonstrated. The bottleneck for single molecule SC in the vibrational case is the fabrication of low-mode volume cavities in the infrared (IR) regime~\cite{Nitzan_ir_2022}. This calls for alternate strategies to attain vibrational SC with a single or few molecules.\\

In this work, we propose using optomechanics as a way to achieve SC for molecular vibrations. Over the last decade, optomechanics has emerged as a powerful tool in quantum technologies with applications~\cite{opt_mech_rev_2014} such as backaction cooling of a mechanical oscillator~\cite{Clerk1_2019, Metzger2004}, parametric amplification~\cite{Massel2011, Clerk2_2010}, optomechanically induced transparency~\cite{Fleischhauer2005, Safavi-Naeini2011}, and generation of non-classical quantum states~\cite{Hendrik_2016, Myestre2013}. Aspelmeyer and co-workers have demonstrated SC in an optomechanical architecture, for a micromechanical resonator coupled to an optical cavity setup~\cite {Groblacher2009}. It has been shown recently that surface-enhanced and cavity-enhanced Raman spectroscopy (SERS/CERS) can be understood through the theoretical framework of cavity optomechanics~\cite{Roelli2016, Baumberg_2017, Esteban1_2020, Javier1_2019, Javier2_2016, Esteban2_2022}. Here we exploit this observation to demonstrate that a single molecule in a CERS setup, under strong illumination of a red-detuned laser can be a viable platform to achieve the long-standing goal of single and few-molecule vibrational polaritons. Few-molecule polaritons do not suffer from the deleterious effects of a macroscopic number of dark states, and hence are better candidates for harnessing the properties of polaritons~\cite{raphaelrev2018}.\\

As a proof-of-concept application of few-molecule vibrational polaritons, we will introduce the intriguing concept of coupling chemical reactions via the latter. Biological systems use coupled chemical reactions and thermodynamics to their advantage by driving energetically uphill processes, such as  active transport, using spontaneous reactions, like the dissociation of ATP~\cite{tinoco2013physical}. Humans have looked towards nature for inspiration and translated biological knowledge into innovative products and processes~\cite{breslow2009biomimetic, hong2010biomimetic}. We shall show how the delocalization of the polariton modes inside the cavity can be exploited to design a biomimetic of ATP-driven molecular machines. 
\section*{Results and Discussion}
\subsection*{Model}
\begin{figure*}[htbp]
	\includegraphics[width=\linewidth]{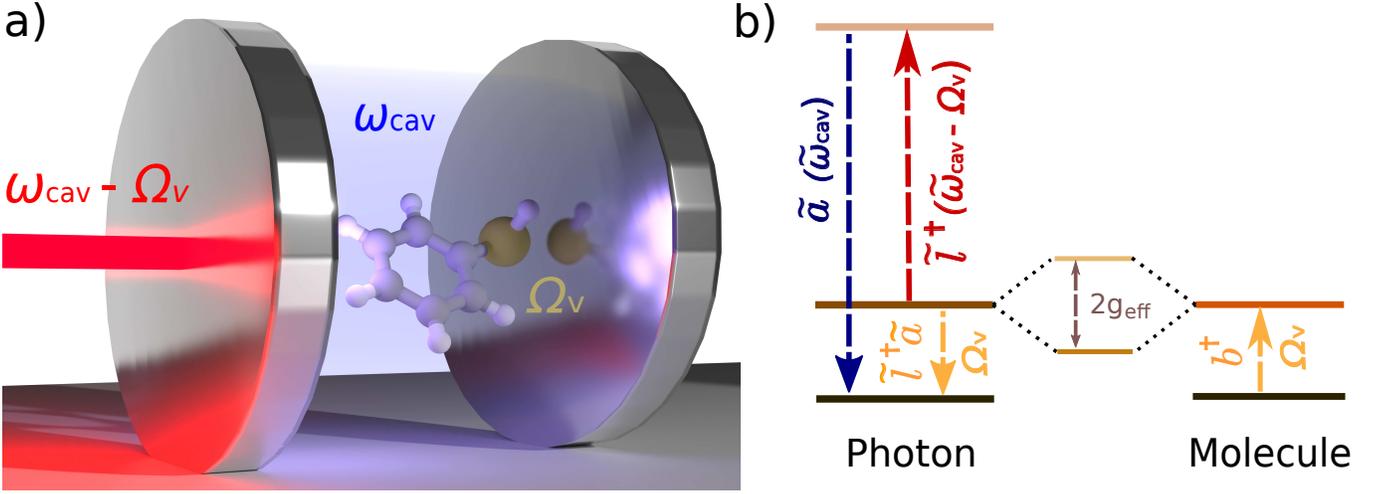} 
	\caption{Experimental setup for achieving vibrational SC in cavity-enhanced Raman spectroscopy (CERS). a)  A single molecule is placed inside a UV-vis cavity ($\omega_{\text{cav}}$) detuned from any electronic transition of the molecule. The cavity is illuminated with a laser red-detuned from the cavity ($\omega_{\text{cav}}-\Omega_{\text{v}}$), where $\Omega_{\text{v}}$ is the frequency of the molecular vibration of interest. b) Schematic depicting the coupling of the composite photon mode denoted by annihilation operator $\tilde{l}^\dagger \tilde{a}$, where $\tilde{l}$ denotes the `laser-like' and $\tilde{a}$ denotes the `cavity-like' normal mode of the laser-cavity subsystem; the molecular vibration denoted with annihilation operator $b$. The composite photon mode and the molecular vibration strongly couple with effective coupling $g_{\text{eff}}$, to give the polaritons. Here the laser-cavity detuning, 
 $\Delta=\Omega_{\text{v}}$ and the cavity-laser coupling, $J$ $\ll \Delta$.}
	\label{fig:cav} 
\end{figure*}

Our theoretical model considers a single molecule placed inside a UV-vis cavity, such that the cavity frequency is off-resonant to any optically allowed transitions of the molecule. In this regime, the coupling between the cavity and the molecule is purely parametric through the molecule's polarizability, with the vibration of the molecule causing a dispersive shift in the cavity resonance~\cite{Roelli2016}. Due to better spatial overlap between the mode profile of the cavity and the molecule, we consider using a Fabry-Per\'ot cavity. However, the formalism presented here is valid for other cavity types. We show that the cavity-molecule system, when pumped with a laser that is red-detuned from the cavity and the detuning is in the order of the molecule's vibrational frequency 
 (Figure~\ref{fig:cav}b), yields an effective Hamiltonian resembling the vibrational polaritonic Hamiltonian. Importantly, the light-matter coupling can be tuned by varying the laser power.\\

We model the photon mode and the vibration of the molecule as  harmonic oscillators with frequencies $\omega_{\text{cav}}$ and $\Omega_{\text{v}}$, and annihilation operators $a$ and $b$, respectively. The cavity has a decay rate of $\kappa$ and the vibrational mode has a decay rate of $\gamma$. In this work, the losses will be modeled using Lindblad master equations~\cite{nitzan2013chemical}. Since the polarizability, $\alpha$, of the molecule, to the leading order, depends on its vibrational displacement~\cite{Javier2_2016}, $x_v={x_{\text{zpf},v}} (b^\dagger + b)$, the Hamiltonian for the cavity-molecule system is~\cite{Roelli2016}
\begin{equation}
    H_{\text{C-M}}= \hbar [\omega_{\text{cav}} + {g_0}  (b^\dagger + b)] a^\dagger a + \hbar \Omega_{\text{v}} b^\dagger b, 
\end{equation}
where $g_0 = {x_{\text{zpf},v}}\bigg(\omega_{\text{cav}} \frac{\partial \alpha}{\partial x_v} \frac{1}{\epsilon_0 V_\text{c}}\bigg)$ is the vacuum cavity-molecule coupling,  with $V_\text{c}$ and $\epsilon_0$ as the mode volume of the cavity and vacuum permittivity, respectively. This Hamiltonian formally resembles an optomechanical setup~\cite{Roelli2016}, where the displacement of a mechanical oscillator modulates the frequency of the cavity. The cavity then acts back on the oscillator through radiation pressure force, which is a function of the  cavity's photon occupation.\\

We drive the cavity with a laser red-detuned $(\omega_{\text{L}}=\omega_{\text{cav}}-\Delta$, $\Delta>0)$ from the cavity resonance. The full Hamiltonian for a laser mode with annihilation operator $l$ coupled to the cavity-molecule subsystem is given as
\begin{eqnarray}
   H_{\text{full}} &=& H_{\text{C-M}} +\hbar \omega_{\text{L}} l^\dagger l +\hbar J (l^\dagger + l) (a^{\dagger}+a),
\end{eqnarray}
with the cavity-laser coupling $J = \sqrt{\frac{\kappa}{\tau_{\text{rt}_\text{L}}}}$. Here,  $\tau_{\text{rt}_\text{L}}$ is related to the laser power $P=\frac{n_\text{L}\hbar\omega_{\text{L}}} {\tau_{\text{rt}_\text{L}}}$ with $n_\text{L}=\langle l^\dagger l\rangle$ being the mean photon number in the laser mode~\cite{steck2007quantum}.
\\

We make the rotating wave approximation (RWA) in the laser-cavity coupling and diagonalize the laser-cavity subsystem. The operators $\tilde{l}$ and $\tilde{a}$ represent the new `laser-like' and `cavity-like' normal modes of the laser-cavity subsystem with frequencies $\tilde{\omega}_{\text{L}}$ and $\tilde{\omega}_{\text{cav}}$, respectively. The Hamiltonian after making the approximation and change of basis is

\begin{eqnarray}
\nonumber H^{\text{RWA}}_{\text{full}}&=&\hbar\Omega_{\text{v}}b^{\dagger}b+\hbar\tilde{\omega}_{\text{cav}}\tilde{a}^{\dagger}\tilde{a}+\hbar\tilde{\omega}_{\text{L}}\tilde{l}^{\dagger}\tilde{l}+\hbar g_{0}\big(\cos\varphi\cdot\tilde{a}\\
& &+\sin\varphi\cdot\tilde{l}\big)^{\dagger}\big(\cos\varphi\cdot\tilde{a}+\sin\varphi\cdot\tilde{l}\big)(b^{\dagger}+b),
\end{eqnarray}
where $\varphi=\frac{1}{2}\tan^{-1}\bigg(\frac{2J}{\omega_{\text{cav}}-\omega_{\text{L}}}\bigg)$ is the laser-cavity mixing angle~\cite{mandal2020polariton}. \\

Considering the red-detuned case for the cavity-laser detuning, $\Delta$, in the order of ${O}(\Omega_{\text{v}})$, we drop the off-resonant contributions in the cavity-molecule interaction term, simplifying $H_{\text{full}}$ to
\begin{eqnarray}   
 \nonumber H_{\text{R}} &=& \hbar\Omega_{\text{v}}b^{\dagger}b + \hbar\tilde{\omega}_{\text{cav}}\tilde{a}^{\dagger}\tilde{a}+\hbar\tilde{\omega}_{\text{L}}\tilde{l}^{\dagger}\tilde{l} + \frac{\hbar g_{0}}{2}\sin(2\varphi) \cdot \\ & & (\tilde{l}{}^{\dagger}\tilde{a}b^{\dagger}+\tilde{l}\tilde{a}^{\dagger}b).
\end{eqnarray}
We will later set the laser-cavity detuning $\Delta=\Omega_{\text{v}}$.\\

We define a composite laser-cavity photon mode with annihilation operator $\mathcal{A}_{\text{ph}}= \frac{\tilde{l}^{\dagger}\tilde{a}}{\sqrt{(\tilde{n}_\text{L}-\tilde{n}_a)}}$ (Figure~\ref{fig:cav}b), where $\tilde{n}_{\text{L}}=\langle \tilde{l}^\dagger 
\tilde{l}\rangle$ and $\tilde{n}_a=\langle \tilde{a}^\dagger \tilde{a}\rangle$ are the mean photon occupations in the `laser-like' and `cavity-like' normal modes, respectively. The Heisenberg equations of motion for the operators $\mathcal{A}_{\text{ph}}$ and $b$ in the mean-field approximation~\cite{F.Ribeiro2018}, 

\begin{subequations}
\begin{align}
\frac{d}{dt}\mathcal{A}_{\text{ph}} &= -i(\tilde{\omega}_{\text{cav}}-\tilde{\omega}_{\text{L}})\mathcal{A}_{\text{ph}}-\frac{ig_{0}}{2}\sin(2\varphi) \sqrt{\tilde{n}_{\text{L}}-{\tilde{n}_{a}}} \cdot b,&\\
  \frac{d}{dt}b &= -i\Omega_{\text{v}}b-\frac{ig_{0}}{2}\sin(2\varphi) \sqrt{\tilde{n}_{\text{L}}-{\tilde{n}_{a}}}\cdot\mathcal{A}_{\text{ph}},&
\end{align}
\end{subequations}
yield the effective Hamiltonian
\begin{eqnarray}
   \nonumber H_{\text{eff}} &\approx& \hbar\omega_{\text{ph}}\mathcal{A}_{\text{ph}}^{\dagger}\mathcal{A}_{\text{ph}}+\hbar\Omega_{\text{v}}b^{\dagger}b+\hbar g_{0}\bigg( \frac{J}{\Delta} \bigg) \sqrt{n_{\text{L}}}\cdot\\
    & & \big(\mathcal{A}_{\text{ph}}^{\dagger}b+\mathcal{A}_{\text{ph}}b^{\dagger}\big),
\end{eqnarray}
when ${n}_\text{L}\gg\tilde{n}_{a}$ and $J\ll\Delta$ (see Supplementary note-1). Here we have suggestively defined $\omega_{\text{ph}}\equiv\Delta$. For an input laser drive with power $P$, the interaction strength between the composite photon and the molecular vibration transforms to  $g_{\text{eff}}=\frac{g_0}{\Delta} \sqrt{\frac{P\kappa}{\hbar \omega_{\text{L}}}}$, consistent with the results obtained from the classical treatments of the laser mode~\cite{Esteban2_2022, Javier1_2019, opt_mech_rev_2014, steck2007quantum}. \\

\begin{figure*}[htbp]
	\includegraphics[width=\linewidth]{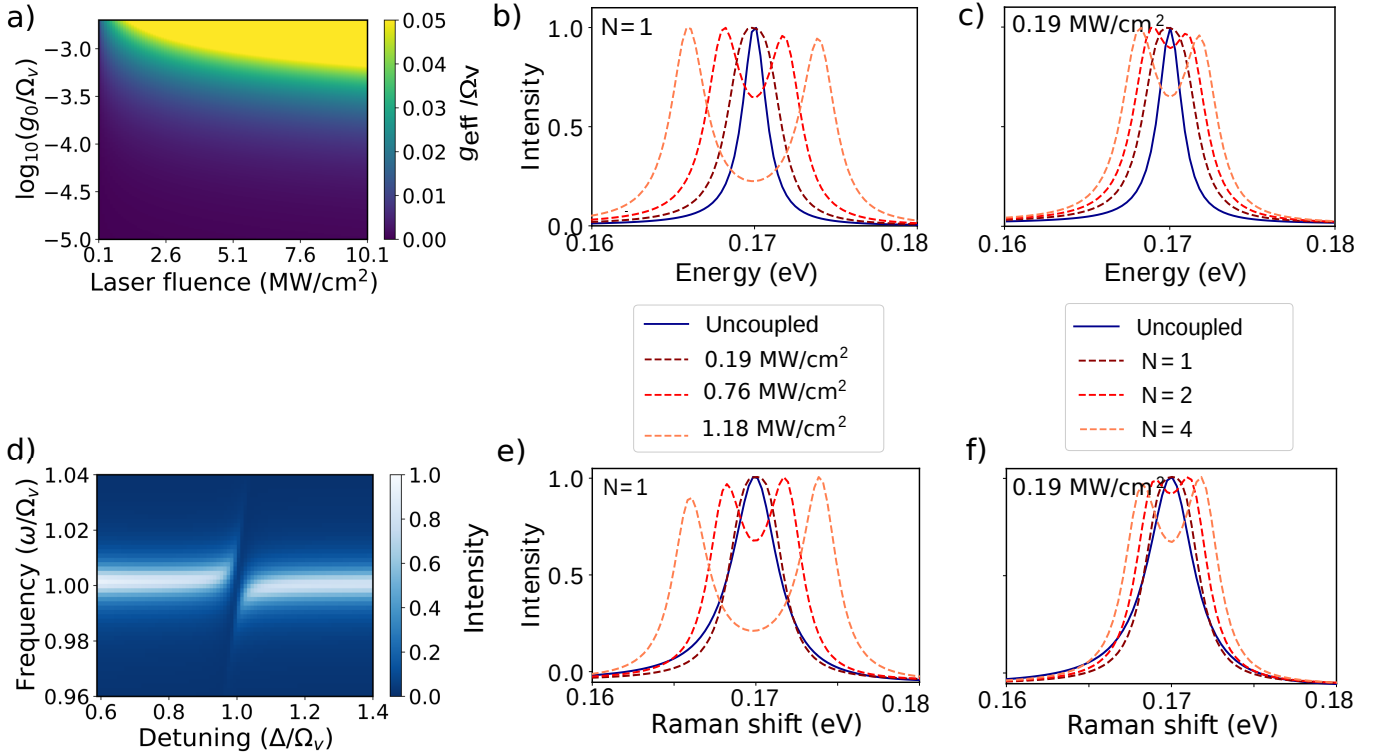} 
	\caption{Spectroscopic signatures of few-molecule vibrational strong coupling in CERS setup. a) The effective light-matter coupling ($g_{\text{eff}}$) as a function of the single-photon coupling ($g_0$) and laser fluence ($P/A$). Vibrational spectra of the molecule as a function of b) laser fluence, c) number of molecules ($N$), d) Cavity-laser detuning ($\omega_{\text{cav}}-\omega_{\text{L}}= \Delta$). Emission spectra from the cavity as a function of e) laser fluence, f) number of molecules,  for  $\hbar\Omega_{\text{v}}=0.17$ eV, $\omega_{\text{cav}}=1.7$ eV, $\omega_{\text{ph}}=\Omega_{\text{v}}$, $g_0=1.5\times10^{-3}\Omega_{\text{v}}$, $\kappa=0.02\Omega_{\text{v}}$, $\gamma=\gamma_r/2 + \gamma_{\text{pd}}= 0.01\Omega_{\text{v}}$, $\gamma_\text{r}/2=10^{-4}\Omega_{\text{v}}$, and $A=5$ $\mu$m$^2$ unless otherwise mentioned.}  
	\label{fig:spectra} 
\end{figure*}

When $\omega_{\text{ph}} = \Omega_{\text{v}}$, $H_\text{eff}$ resembles a vibrational polaritonic Hamiltonian, where the composite photon mode is resonant with the vibrational DOF (Figure~\ref{fig:cav}b)~\cite{raphaelrev2018}. Here the coupling strength is tunable by changing the pumping power of the laser. This can, in principle, foster the SC regime when the coupling strength supersedes the decay processes in the system. To look at parameter sets yielding this regime and to compute spectra, we simulate the dynamics of the density matrix ($\rho$) of the system using Lindblad master equations~\cite{Javier2_2016,nitzan2013chemical} given as
\begin{align}
  \nonumber  \frac{\partial \rho}{\partial t} =& i [H_\text{eff},\rho] + \frac{\kappa}{2}\mathcal{L}_\mathcal{A} [\rho] + \frac{(n^{\text{th}}_v+1)\gamma_r}{2}\mathcal{L}_b[\rho] + &\\  & \frac{n^{\text{th}}_v\gamma_r}{2}\mathcal{L}_{b^\dagger}[\rho] + {n^{\text{th}}_v\gamma_{\text{pd}}} \mathcal{L}_{b^\dagger b}[\rho].&
\end{align}
The last four terms on the right-hand side are the Lindblad-Kossakowski terms defined as $\mathcal{L}_\mathcal{O}[\rho]=2\mathcal{O}\rho\mathcal{O}^\dagger - [\mathcal{O}^\dagger\mathcal{O},\rho]$. Here, $\mathcal{L}_\mathcal{A}$ models the incoherent decay from the composite photon mode. The incoherent decay, thermal pumping, and pure dephasing of the vibrational mode by the environment at temperature $T$ are modeled by 
the $\mathcal{L}_b$, $\mathcal{L}_{b^\dagger}$, and $\mathcal{L}_{b^\dagger b}$ terms, respectively, where $n^{\text{th}}_v=(e^{\hbar\Omega_{\text{v}}/k_{\text{B}}T} -1)^{-1}$ is the Bose-Einstein distribution function at transition  energy $\hbar\Omega_{\text{v}}$. Additionally, in the limit of large photon number in the laser ($n_{\text{L}}\gg 1$), assuming the photon occupation to be constant, and thus the laser mode to be non-lossy, the decay rate of the composite photon equals the cavity decay rate $\kappa$ (see Supplementary note-2). \\ 
The simulations have been performed using the QuTip package~\cite{qutip1, qutip2} and the results are presented in Figure~\ref{fig:spectra} for the molecule Rhodamine 6G~\cite{Roelli2016}, where $\hbar\Omega_{\text{v}}=0.17$ eV, $\omega_{\text{ph}}= \Omega_{\text{v}}$ ($0.17$ eV), $\gamma= \gamma_\text{r}/2 + \gamma_{\text{pd}}= 0.01\Omega_{\text{v}}$ ($1.7\times10^{-3}$ eV), $\kappa=0.02\Omega_{\text{v}}$ ($3.4\times10^{-3}$ eV), $g_0= 1.5\times10^{-3}\Omega_{\text{v}}$ ($2.6\times10^{-4}$ eV). Here, $\gamma_\text{r} = 10^{-4}\Omega_{\text{v}}$ ($1.7\times10^{-5}$ eV) and  $\gamma_{\text{pd}}$ are the rates for vibrational relaxation and pure dephasing, respectively~\cite{mattnoneq2021} . The fluence of the lasers is  chosen to be below $\sim 10$ MW$/\text{cm}^2$~\cite{Roelli2016, Werner2011} with a beam area of $A= 5$ $\mu$m$^2$. Figure~\ref{fig:spectra}a shows the effective light-matter coupling, $g_{\text{eff}}$, as a function of laser fluence, $P/A$, and single photon coupling strength, $g_0$. In Figure~\ref{fig:spectra}b, the vibrational spectrum of the molecule $S_b(\omega)=\text{Re}[\int_{0}^{\infty} e^{-i\omega t} \langle x_{\text{v}}(t)x_{\text{v}}(0) \rangle_{\text{ss}}\text{d}t]$, splits, demonstrating SC. We see the Rabi-splitting increases with laser power, thus giving us additional control over the light-matter coupling strength. Figure~\ref{fig:spectra}c shows the spectra with one, two, and four molecules for constant laser power. Figure~\ref{fig:spectra}d is the vibrational spectrum of the molecule as a function of the cavity-laser detuning. The avoided crossing at the detuning ($\Delta$) equal to the vibrational frequency ($\Omega_{\text{v}}$) demonstrates maximal hybridization between the photonic and matter DOF. Finally, Figure ~\ref{fig:spectra}e and f show the emission spectra from the cavity at steady state (\text{ss}), $S_{\mathcal{A}}(\omega)=  \omega^4 \cdot \text{Re}[\int_{0}^{\infty}e^{-i\omega t} \langle \mathcal{A}_{\text{ph}}^\dagger(t)\mathcal{A}_{\text{ph}}(0)\rangle_{\text{ss}} \text{d}t]$~\cite{spectra_3, Javier2_2016, Dezfouli2019}, also revealing the polariton peaks. 

\subsection*{Polariton-assisted thermodynamic driving}
\begin{figure*}[htbp]
	\includegraphics[width=\linewidth]{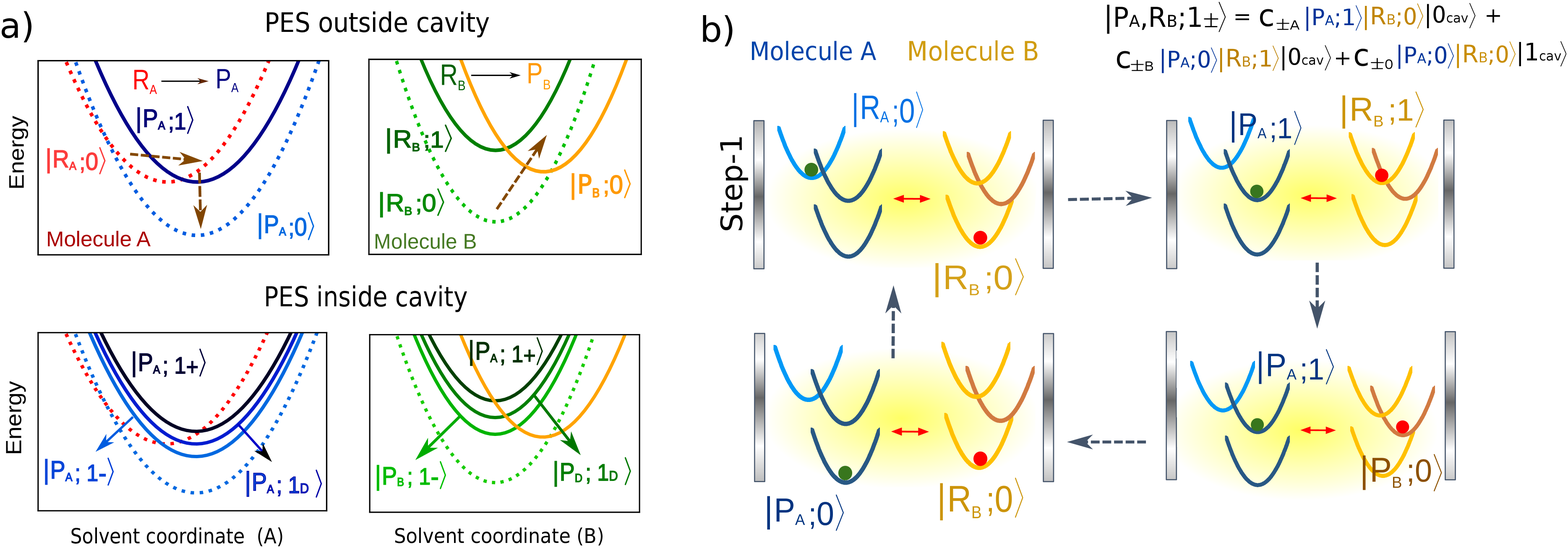} 
	\caption{a) Schematic PES for molecule A and B, outside and inside the cavity. Both A and B undergo electron transfer reactions with negative and positive $\Delta G$, respectively. The dashed arrows show the main reaction pathways for each molecule. The direction (upward or downward) of the arrow indicates whether the reaction is uphill or downhill and the steepness indicates the transition energy. The reaction takes molecule A from $|\text{R}_\text{A}\rangle$ to $|\text{P}_\text{A}\rangle$,  and molecule B  from 
 $|\text{R}_\text{B}\rangle$ to $|\text{P}_\text{B}\rangle$. The PES are labeled as $|\text{E},v_{\text{E}}\rangle$, where $\text{E}$ labels the electronic state, R$_i$ or P$_i$ for $i\in\{\text{A, B}\}$. Outside the cavity, $|v_{\text{E}}\rangle$ represents the vibrational state of the high-frequency mode corresponding to the electronic state $|\text{E}\rangle$. The coupling of this high-frequency vibrational mode to the composite photon mode leads to the two polariton states, $|1_{\pm}\rangle$, and one dark state, $|1_{\text{D}}\rangle$. b) 
 One cycle of the coupled reactions. \emph{Step-1}: We start from the reactant electronic states and vibrational ground states in both molecules ($|\text{R}_\text{A};0\rangle$, $|\text{R}_\text{B};0\rangle$). \emph{Step-2}: As molecule A reacts spontaneously ($|\text{R}_\text{A};0\rangle \rightarrow |\text{P}_\text{A};1\rangle$), the polariton modes, $|\text{P}_{\text{A}},\text{R}_{\text{B}};1_{\pm}\rangle$, being delocalized also promote vibrational excitation in B from $|\text{R}_\text{B};0\rangle$ to $|\text{R}_\text{B};1\rangle$. \emph{Step-3}: This allows B to react  from its excited state ($|\text{R}_{\text{B}};1\rangle\rightarrow|\text{P}_\text{B};0\rangle$). \emph{Step-4}: Finally, both A and B relax to $|\text{P}_\text{A};0\rangle$ and $|\text{R}_\text{B};0\rangle$, respectively, after which the molecule A needs to be replaced for the next cycle (PES have been drawn not to scale, to emphasize the mechanism.)}
	\label{fig:schematic} 
\end{figure*}

The matter component of the polariton modes is delocalized over many molecules under collective SC~\cite{dunkelberger2022vibration,raphaelrev2018}. This delocalization can be exploited more effectively with few-molecule polaritons, owing to reduced involvement of dark modes~\cite{Chikkaraddy2016, mattdark2022}, which remain parked essentially at the same energy as the original molecular transitions. In this work, we consider the molecular species undergoing electron-transfer reactions, modeled using Marcus-Levich-Jortner (MLJ) theory~\cite{marcus1_1964, levich1966, jortner_1976}. Our system consists of two reactive molecules A and B of different species placed inside an optomechanical cavity. Here  molecule A features a spontaneous reaction (with negative free energy change, $\Delta G_{\text{A}}<0$), while molecule B features an endergonic reaction ($\Delta G_{\text{B}}>0$, with $\Delta G_{\text{B}} > k_{\text{B}} T$). We demonstrate thermodynamic coupling between the two molecular species via the composite photon mode, such that the spontaneous electron transfer in A can drive B to react. Schematically, electron transfer in A creates a vibrationally hot product (Figure~\ref{fig:schematic}a), which,  outside the cavity, just decays to the product ground state. However, inside the cavity, in the timescale of the Rabi frequency, this excitation can be captured by the photon mode, which then can excite the reactant in B to its vibrational excited state. The electron transfer in B can then proceed spontaneously from the reactant's excited state (Figure~\ref{fig:schematic}b). Notably, this scheme can also be generalized to other types of reactions.\\

Within the framework of MLJ theory, the molecules can exist in either of the two diabatic electronic states: $|\text{R}_i\rangle$ corresponding to the reactant and $|\text{P}_i\rangle$ corresponding to the product for molecule  $i\in\{\text{A}, \text{B}\}$ (Figure~\ref{fig:schematic}a). For molecule B, in this case the switching between $|\text{R}_\text{B}\rangle$ and $|\text{P}_\text{B}\rangle$ through electron transfer contributes to useful mechanical work, manifested in changes of nuclear configuration~\cite{pcet1_2015,pcet2_1981,pcet3_1981}. The electronic states for each molecule are dressed with a local high-frequency intramolecular vibrational coordinate represented by annihilation operator $a_{x, i}$, for $x\in\{\text{R}, \text{P}\}$, and coupled to a low-frequency effective solvent mode treated classically with rescaled momentum and position as $p_{\text{S}, i}$ and $q_{\text{S}, i}$. respectively. 
We assume that the high-frequency modes of both species, being resonant, are the only ones that couple to the 
 composite photon~\cite{jorge2019}. Upon reaction, the high-frequency mode undergoes a change in its equilibrium configuration according to, $a_{\text{R}, i} = D^\dagger_i a_{\text{P}, i} D_i$, where $D_i=\text{exp}[(a^\dagger_{\text{P}, i} - a_{\text{P}, i})\sqrt{S_i}]$ is the displacement operator, and $S_i$ is the Huang-Rhys factor~\cite{sindhana2022}. The Hamiltonian describing the system is given as $H = H_0 + V_{\text{react}}$, where
 \begin{subequations}
\begin{align}
   H_0 =& H_{\text{ph}}+ \sum_{i=\text{A},\text{B}} \sum_{x=\text{R},\text{P}} (H_{x,i} + V_{x,i})|x_i\rangle\langle x_i|,& \\
     V_{\text{react}} =& \sum_{i=\text{A,B}} J_i (|\text{R}_i\rangle\langle \text{P}_i| + |\text{P}_i\rangle\langle \text{R}_i|).&
\end{align}
\end{subequations}
Here $H_{\text{ph}}=\hbar \omega_{\text{ph}} \big(\mathcal{A}_{\text{ph}}^\dagger \mathcal{A}_{\text{ph}}+\frac{1}{2}\big)$ is the bare Hamiltonian corresponding to the composite photon mode consisting of the laser and the cavity, $H_{x, i}$ represents the high-frequency mode and the solvent mode associated with molecule $x_i$,
\begin{subequations}
\begin{align}
     H_{\text{R},i} =& \hbar \Omega_{\text{R},i}\bigg(a^\dagger_{\text{R}, i}a_{\text{R}, i}+\frac{1}{2}\bigg) + \frac{1}{2}\hbar\Omega_{\text{S},i}(|p_{\text{S}, i}|^2 + |q_{\text{S}, i}|^2),&\\
      \nonumber H_{\text{P},i} =& \hbar \Omega_{\text{P},i}\bigg(a^\dagger_{\text{P}, i}a_{\text{P}, i}+\frac{1}{2}\bigg) + \frac{1}{2}\hbar\Omega_{\text{S},i}(|p_{\text{S}, i}|^2 + & \\ 
      & |q_{\text{S}, i}+ d_{\text{S},i}|^2)  +\Delta G_i,&
\label{pot_mar}
\end{align}
\end{subequations}
with $d_{\text{S},i}$ and $\Omega_{\text{S},i}$ being the displacement and frequency along the solvent coordinate, respectively, and $\Delta G_i$,  the free energy difference for the molecular species $i$. Additionally, $V_{x,i}=\hbar g_{x,i}(a_{x, i}\mathcal{A}_{\text{ph}}^\dagger + a^\dagger_{x, i}\mathcal{A}_{\text{ph}})$ is the effective coupling between the photonic and molecular DOF. For simplicity, we assume that the reaction involves a vibrational mode with nearly identical frequency and light-matter coupling strength for species A and B in both the reactant and product electronic states ($\Omega_{x, i}=\Omega_{y,j}\equiv\Omega_{\text{v}}$ and $g_{x, i}=g_{y,j}\equiv g$). Finally, the diabatic couplings between the electronic states $|\text{R}_i\rangle$ and $|\text{P}_i\rangle$ are given by $V_{\text{react}}$, where $ J_{i}$ is the coupling strength.\\

\begin{figure*}[htbp]
	\includegraphics[width=1\linewidth]{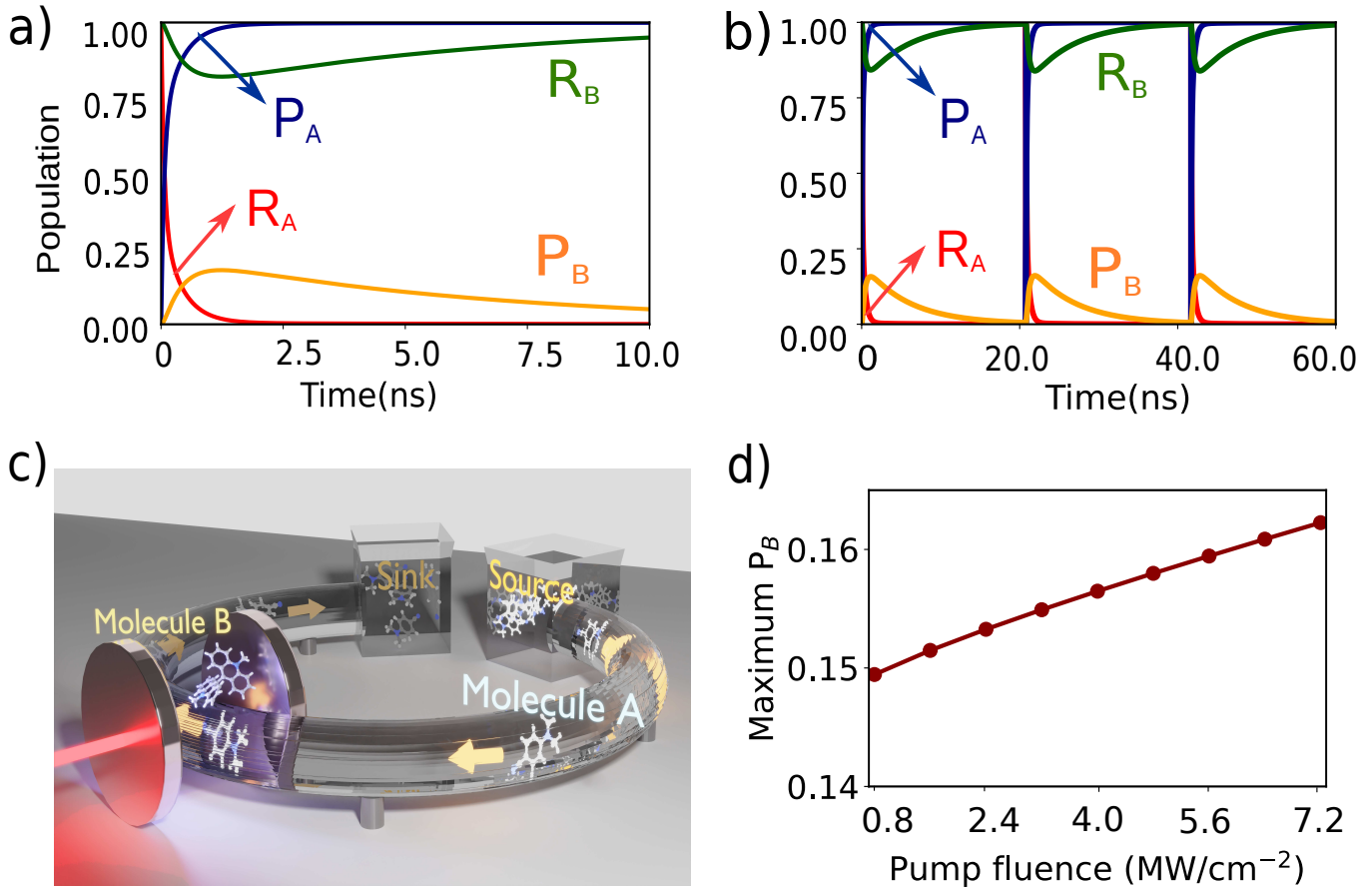} 
	\caption{Simulations showing polariton-assisted thermodynamic driving. Population dynamics of molecule A and B in a) short times and b) long times, with molecule A being replenished through the flow apparatus depicted in c). Molecule B switches from $|\text{R}_\text{B}\rangle$ to $|\text{P}_\text{B}\rangle$ in a cycle, thus producing mechanical work. c)  Proposed experimental setup for continuous mechanical work from molecule B by the circulation of molecule A. Using a flow chemistry apparatus, molecule A in state $|\text{R}_\text{A};0\rangle$ state (glowing) is transported from the `source' bath to the cavity, where it transforms to $|\text{P}_\text{A};0\rangle$ (no glow), drives molecule B from $|\text{R}_\text{B};0\rangle$ to $|\text{P}_\text{B};0\rangle$, and subsequently flow to the `sink' bath. d) Maximum population reached in P$_\text{B}$ as a function of the fluence of the laser drive. Here $\hbar\Omega_{\text{R},i}=\hbar\Omega_{\text{P},i}=0.22$ eV (call $\hbar\Omega_{\text{v}}$),  $g_0 =2\times10^{-3}\Omega_{\text{v}}$, $P/A=6.4$ MW/cm$^2$, A=5 $\mu$m$^2$, $\gamma_\text{A}=\gamma_\text {B}=1 \times 10^{-5} \Omega_{\text{v}}$, $\omega_{\text{cav}}=2.2$ eV, $\kappa=0.015 \Omega_{\text{v}}$, $\lambda_{\text{S}}^{\text{A}}=0.04\hbar\Omega_{\text{v}}$, $\lambda_{\text{S}}^{\text{B}}=0.1\hbar\Omega_{\text{v}}$, $\Delta G_\text{A}=-\hbar\Omega_{\text{v}}$, $\Delta G_\text{B}=0.7\hbar\Omega_{\text{v}}$, $\hbar\Omega_{\text{cut}}=0.1\hbar\Omega_{\text{v}}$, and $\eta=0.0001$ unless otherwise specified.   }
 	\label{fig:population} 
\end{figure*}

We can solve $H_0$ parametrically as a function of the solvent coordinates to construct the potential energy surfaces(PES) (Figure ~\ref{fig:population}a and b). Considering these diabatic couplings $V_{\text{react}}$ to be perturbative, $H_0$ can be diagonalized to obtain the two polariton modes, $a^{(\pm)}_{x_\text{A},y_\text{B}}$, with frequencies $\omega_{\pm}=\frac{1}{2}(\omega_{\text{ph}}+ \Omega_{\text{v}} \pm \sqrt{(\omega_{\text{ph}} + \Omega_{\text{v}})^2 + 8g^2} )$, and one dark mode, $a^D_{x_\text{A},y_\text{B}}$, with frequency $\Omega_{\text{v}}$, given as
\begin{subequations}
\begin{eqnarray}
a^{(\pm)}_{x_\text{A},y_\text{B}} &=& \cos{\theta} \mathcal{A}_{\text{ph}} \pm \sin{\theta} \cdot \frac{1}{\sqrt{2}}(a_{x,\text{A}} + a_{y,\text{B}}), \\
    a^D_{x_\text{A},y_\text{B}} &=& c_{x,\text{A}}a_{x,\text{A}} + c_{y,\text{B}}a_{y,\text{B}},
\end{eqnarray}
\end{subequations}
such that, $c_{x, \text{A}}+c_{y, \text{B}}=0$ and $|c_{x, \text{A}}|^2 + |c_{y, \text{B}}|^2=1$. 
Here, $\theta = \frac{1}{2}\tan^{-1}\bigg(\frac{2\sqrt{2}g}{\delta}\bigg)$ is the mixing angle, where $\delta= (\omega_{\text{ph}}-\Omega_{\text{v}})$ is the detuning between the composite photon mode and the molecule. Here we have chosen the composite photon mode to be resonant with the intramolecular vibration, \emph{i.e.}, $\delta=0$.\\

We now define multi-particle states $|\bm{\phi};\bm{\nu}_{\bm{\phi}}\rangle$ that span the Hilbert space of the system, where $|\bm{\phi}\rangle=|x_\text{A},y_\text{B}\rangle$, $x,y\in\{\text{R}, \text{P}\}$ corresponds to the electronic DOF, and $|\bm{\nu}_{\bm{\phi}}\rangle=|\nu^{+}_{x_\text{A},y_\text{B}},\nu^{-}_{x_\text{A},y_\text{B}},\nu^D_{x_\text{A},y_\text{B}}\rangle$ to the cavity-vibrational mode of each electronic state $|\bm{\phi}\rangle$~\cite{mattnoneq2021}. To describe the reaction, we look at the population dynamics in the electronic states. The kinetic master equations governing the time evolution of the system are given as~\cite{mattnoneq2021}
\begin{align}
    \nonumber \frac{d p_{(\bm{\phi};\bm{\nu}_{\bm{\phi}})}(t)}{dt} =& -\bigg[\sum_{(\bm{\phi}';\bm{\nu}'_{\bm{\phi}'})\neq(\bm{\phi};\bm{\nu}_{\bm{\phi}})} k(\bm{\phi}';\bm{\nu}'_{\bm{\phi}'}|\bm{\phi};\bm{\nu}_{\bm{\phi}})\bigg] p_{(\bm{\phi};\bm{\nu}_{\bm{\phi}})}&\\
     & +  \sum_{(\bm{\phi}';\bm{\nu}'_{\bm{\phi}'})\neq(\bm{\phi};\bm{\nu}_{\bm{\phi}})} k(\bm{\phi};\bm{\nu}_{\bm{\phi}}|\bm{\phi}';\bm{\nu}'_{\bm{\phi}'})\cdot p_{(\bm{\phi'};\bm{\nu'}_{\bm{\phi}'})},&
     \label{kinetic}
\end{align}
where $p_{(\bm{\phi};\bm{\nu}_{\bm{\phi}})}(t)$ represents the population in $|\bm{\phi};\bm{\nu}_{\bm{\phi}}\rangle$ state, and $k(\bm{\phi}';\bm{\nu}'_{\bm{\phi}'}|\bm{\phi};\bm{\nu}_{\bm{\phi}})$ is the rate constant for population transfer from $|\bm{\phi};\bm{\nu}_{\bm{\phi}}\rangle$ to $|\bm{\phi}';\bm{\nu}'_{\bm{\phi}'}\rangle$ due to processes like reactive transitions between the electronic states accompanied by solvent reorganization and decay through the cavity and vibrational DOF. The rate constant for the reactive transition at a temperature $T$ within the framework of MLJ theory is given as~\cite{jorge2019}
\begin{align}
    \nonumber k(\bm{\phi}';\bm{\nu}'_{\bm{\phi}'}|\bm{\phi};\bm{\nu}_{\bm{\phi}}) =& \sqrt{\frac{\pi}{\lambda_{\text{S}}^{(\bm{\phi}\bm{\phi'})}k_{\text{B}}T}}\frac{|J_{\bm{\phi}\bm{\phi'}}|^2}{\hbar} |\langle \bm{\nu'}_{\bm{\phi}'}|\bm{\nu}_{\bm{\phi}}\rangle|^2 &\\ &
     \times \exp\bigg[ -\frac{\big(E_{\bm{\phi};\bm{\nu}_{\bm{\phi}}}-E_{\bm{\phi'};\bm{\nu'}_{\bm{\phi}'}}+ \lambda_{\text{S}}^{(\bm{\phi}\bm{\phi'})}\big)^2}{4\lambda_{\text{S}}^{(\bm{\phi}\bm{\phi'})}k_{\text{B}}T}\bigg].&
     \label{marcus}
\end{align}
Here, $E_{\bm{\phi};\bm{\nu}_{\bm{\phi}}} = E_{x_\text{A}}+E_{y_\text{B}}+\hbar\big[\omega_+\big(\nu^+_{x_\text{A}y_\text{B}}+\frac{1}{2}\big) + \omega_-\big(\nu^-_{x_\text{A}y_\text{B}}+\frac{1}{2}\big) + \Omega_{\text{v}}\big(\nu^D_{x_\text{A}y_\text{B}}+\frac{1}{2}\big)\big] $ is the energy of the state $|\bm{\phi};\bm{\nu}_\phi\rangle=|x_\text{A},y_\text{B}\rangle\otimes|\nu^{+}_{x_\text{A},y_\text{B}},\nu^{-}_{x_\text{A},y_\text{B}},\nu^D_{x_\text{A},y_\text{B}}\rangle$, and  $\lambda_{\text{S}}^{(\bm{\phi}\bm{\phi'})}$ and $J_{\bm{\phi}\bm{\phi'}}$ are the solvent reorganization energy and diabatic coupling, respectively, corresponding to the reacting species. Additionally, $\langle \bm{\nu'}_{\bm{\phi}'}|\bm{\nu}_{\bm{\phi}}\rangle$ represent the Franck-Condon factors for the hybrid photon-vibration states $|\bm{\nu'}_{\bm{\phi}'}\rangle$ and $|\bm{\nu}_{\bm{\phi}}\rangle$ corresponding to the electronic states $|\bm{\phi'}\rangle$ and $|\bm{\phi}\rangle$, respectively. For the simulations in this work, the Franck-Condon factors have been computed numerically from eigenstates obtained using 
the standard discrete-variable representation (DVR) of Colbert and Miller~\cite{colbert_1992}.  \\

The reactive transitions transfer populations across different electronic states, while the cavity and vibrational decays lead to dynamics within the same electronic state. In these simulations, since $k_{\text{B}} T \ll \hbar \Omega_{\text{v}}$, we restrict ourselves to the first excitation manifold in the photon-vibration DOF. With the bare vibrational decay rate for the intramolecular vibrations for molecule A and B as $\gamma_\text{A}$ and $\gamma_\text{B}$, respectively, and bare cavity decay rate as $\kappa$, we have 
\begin{equation}
    k(\bm{\phi}, \bm{1}_{q,\bm{\phi}}|\bm{\phi},\bm{0})= |c_{q0}|^2 \kappa + |c_{q\text{A}}|^2 \gamma_\text{A} +  |c_{q\text{B}}|^2 \gamma_\text{B},
\end{equation}
where $\bm{1}_{q,\bm{\phi}}$ represents a single excitation in the polaritons $(q=\pm$) or the dark  ($q=D$) mode, and the $c_{qj}$'s correspond to the expansion coefficients of the excited eigenmode in terms of the cavity and vibrational modes~\cite{hopfield1969}. 

Finally, the anharmonic couplings between the vibrational mode of interest and an other bath of low frequency modes leads to transitions between the polaritons and the dark mode,~\cite{mattnoneq2021}
\begin{align}
  \nonumber k(\bm{\phi}, \bm{1}_{q,\phi}|\bm{\phi},\bm{1}_{q',\phi}) = 2 \pi \bigg( \sum_{i=1}^2 |c_{q'i}|^2|c_{qi}|^2\bigg) \times \{\Theta(-\Omega) \cdot \\ 
   [n^{\text{th}}(-\Omega)+1] \mathcal{J}(-\Omega) + \Theta(\Omega)
    n^{\text{th}}(\Omega) \mathcal{J}(\Omega) \},
\end{align}
where $\Theta(\Omega)$ is the heavyside step function, $n^{\text{th}}(\Omega)$ is the Bose-Einstein distribution function at the transition energy $\hbar\Omega=\hbar (\Omega_{q'}-\Omega_q)$, and $\mathcal{J}(\Omega)$ is the spectral density of the low frequency modes. Assuming the spectral density to be Ohmic~\cite{ohmic_2015}, we have $\mathcal{J}(\Omega)=\eta \Omega \exp[-(\Omega/\Omega_{\text{cut}})^2]$, where $\eta$ is a dimensionless parameter modeling the anharmonic system–bath interactions and $\Omega_{\text{cut}}$ is the cut-off frequency for the low-frequency modes. \\
 
The results of the simulations are presented in Figure~\ref{fig:population}. Here, $\hbar\Omega_{\text{R},i}=\hbar\Omega_{\text{P},i}=0.22$ eV (call $\hbar\Omega_{\text{v}}$),  $g_0 =2\times10^{-3}\Omega_{\text{v}}$ ($4.4\times10^{-4}$ eV), $P/A=6.4$ MW/cm$^2$, A=5 $\mu$m$^2$, $\gamma_\text{A}=\gamma_\text {B}=1 \times 10^{-5} \Omega_{\text{v}}$ ($2.2\times10^{-6}$ eV)~\cite{mattnoneq2021}, $\omega_{\text{cav}}=2.2$ eV, $\kappa=0.015 \Omega_{\text{v}}$ ($3.3\times10^{-3}$ eV), $\lambda_{\text{S}}^{\text{A}}=0.04\hbar\Omega_{\text{v}}$ ($8.8\times10^{-3}$ eV), $\lambda_{\text{S}}^{\text{B}}=0.1\hbar\Omega_{\text{v}}$ ($2.2\times10^{-2}$ eV), $\Delta G_\text{A}=-\hbar\Omega_{\text{v}}$ ($-0.22$ eV), $\Delta G_\text{B}=0.7\hbar\Omega_{\text{v}}$ ($0.15$ eV), $\hbar\Omega_{\text{cut}}=0.1\hbar\Omega_{\text{v}}$ ($2.2\times10^{-2}$ eV), and $\eta=0.0001$~\cite{mattnoneq2021}. The decay rates have been chosen to be similar to those typically found in VSC experiments~\cite{xiang_2018, F.Ribeiro2018, xiang2020intermolecular}. The diabatic couplings are chosen to be $\hbar J_\text{A}=\hbar J_\text{B}=0.005\hbar\Omega_{\text{v}}=(1.1\times10^{-3}$ eV) and $T=298$ K. We start from the initial electronic state $|\text{R}_\text{A}, \text{R}_\text{B}; \bm{0}\rangle$. Independently, the reaction of molecule A is spontaneous  due to its negative free energy change, $\Delta G_\text{A}<0$, while molecule B remains in its thermodynamically stable conformer $|R_\text{B};0\rangle$ (Figure~\ref{fig:population}a and b). This reflects the dynamics of the species outside of the cavity. Placing both the molecules inside the cavity couples the two reactions via the photonic mode enabling the spontaneity of the reaction of molecule A to thermodynamically `lift' B to its unstable configuration $|\text{P}_\text{B};0\rangle$ producing mechanical work. However, after molecule A has fully reacted (change in nuclear configuration), inevitably B has to relax again to its stable configuration $|\text{R}_\text{B};0\rangle$, completing one cycle of the mechanical motion of B (Figure~\ref{fig:schematic}b). The maximum population obtained in $|\text{P}_\text{B}\rangle$ before molecule B relaxes back to $|\text{R}_\text{B}\rangle$ increases with the light-matter coupling strength ($g_{\text{eff}}$), tunable with the fluence of the driving laser (Figure~\ref{fig:population}d). For the cycle to be repeated, molecule A needs to be `recharged' or `replaced'. To achieve this, we envision a flow setup, as schematically depicted in Figure~\ref{fig:population}c, that can circulate molecule A inside and out of the cavity. Continued circulation of the A molecules is essential for the molecular machine of B to be oscillating between reactant and product and producing mechanical work (Figure~\ref{fig:population}a and b). This phenomenon realizes a heat engine producing mechanical work in molecule B, using the (chemical) energy flow from a `source' to a  `sink' bath in the form of molecule A~\cite{morris2012dawn}.  

\section*{Conclusion}
We have shown that the physics of cavity optomechanics can be harnessed in CERS to achieve single to few-molecule vibrational SC using laser-driven UV-vis cavities. We show that the coupling strength and hence the Rabi splitting is tunable with the laser intensity, and it is achievable with realistic pump powers and cavity-molecule couplings. SC in the few molecules regime can avail enhanced polaritonic effects owing to the reduced entropic penalty from the dark states. By using the MLJ theory for electron transfer, we show that the photon-mediated coupling between two reactions, one spontaneous and one non-spontaneous, can be exploited to thermodynamically drive the non-spontaneous process using the spontaneous one. This effect is analogous to harnessing ATP to drive uphill biological processes like the active transport of ions across a membrane against their concentration gradient and can be used to design bio-inspired molecular machines~\cite{Richards2016}. Moving forward, an experimental realization of the scheme for vibrational SC presented here would be a significant step towards utilizing polaritons for chemistry. 




\section*{Code availability}
Computational scripts used to generate the plots in the present article are available by email upon request to the authors.

\begin{acknowledgements}
This work was supported as part of the Center for
Molecular Quantum Transduction (CMQT), an Energy
Frontier Research Center funded by the U.S. Department
of Energy, Office of Science, Basic Energy Sciences under
Award No. DE-SC0021314. A.K. thanks Yong Rui Poh, Kai Schwenickke, Juan B. P\'erez-S\'anchez, Alex Fairhall, and Carlos A. Saavedra Salazar for useful discussions.
\end{acknowledgements}


%


\begin{thebibliography}{81}%
\makeatletter
\providecommand \@ifxundefined [1]{%
 \@ifx{#1\undefined}
}%
\providecommand \@ifnum [1]{%
 \ifnum #1\expandafter \@firstoftwo
 \else \expandafter \@secondoftwo
 \fi
}%
\providecommand \@ifx [1]{%
 \ifx #1\expandafter \@firstoftwo
 \else \expandafter \@secondoftwo
 \fi
}%
\providecommand \natexlab [1]{#1}%
\providecommand \enquote  [1]{``#1''}%
\providecommand \bibnamefont  [1]{#1}%
\providecommand \bibfnamefont [1]{#1}%
\providecommand \citenamefont [1]{#1}%
\providecommand \href@noop [0]{\@secondoftwo}%
\providecommand \href [0]{\begingroup \@sanitize@url \@href}%
\providecommand \@href[1]{\@@startlink{#1}\@@href}%
\providecommand \@@href[1]{\endgroup#1\@@endlink}%
\providecommand \@sanitize@url [0]{\catcode `\\12\catcode `\$12\catcode
  `\&12\catcode `\#12\catcode `\^12\catcode `\_12\catcode `\%12\relax}%
\providecommand \@@startlink[1]{}%
\providecommand \@@endlink[0]{}%
\providecommand \url  [0]{\begingroup\@sanitize@url \@url }%
\providecommand \@url [1]{\endgroup\@href {#1}{\urlprefix }}%
\providecommand \urlprefix  [0]{URL }%
\providecommand \Eprint [0]{\href }%
\providecommand \doibase [0]{https://doi.org/}%
\providecommand \selectlanguage [0]{\@gobble}%
\providecommand \bibinfo  [0]{\@secondoftwo}%
\providecommand \bibfield  [0]{\@secondoftwo}%
\providecommand \translation [1]{[#1]}%
\providecommand \BibitemOpen [0]{}%
\providecommand \bibitemStop [0]{}%
\providecommand \bibitemNoStop [0]{.\EOS\space}%
\providecommand \EOS [0]{\spacefactor3000\relax}%
\providecommand \BibitemShut  [1]{\csname bibitem#1\endcsname}%
\let\auto@bib@innerbib\@empty
\bibitem [{\citenamefont {Kavokin}\ \emph {et~al.}(2017)\citenamefont
  {Kavokin}, \citenamefont {Baumberg}, \citenamefont {Malpuech},\ and\
  \citenamefont {Laussy}}]{microcavitiesbook}%
  \BibitemOpen
  \bibfield  {author} {\bibinfo {author} {\bibfnamefont {A.}~\bibnamefont
  {Kavokin}}, \bibinfo {author} {\bibfnamefont {J.}~\bibnamefont {Baumberg}},
  \bibinfo {author} {\bibfnamefont {G.}~\bibnamefont {Malpuech}},\ and\
  \bibinfo {author} {\bibfnamefont {F.}~\bibnamefont {Laussy}},\ }\href@noop {}
  {\emph {\bibinfo {title} {Microcavities}}},\ Series on Semiconductor Science
  and Technology\ (\bibinfo  {publisher} {OUP Oxford},\ \bibinfo {year}
  {2017})\BibitemShut {NoStop}%
\bibitem [{\citenamefont {Törmä}\ and\ \citenamefont
  {Barnes}(2014)}]{Tormarev_2015}%
  \BibitemOpen
  \bibfield  {author} {\bibinfo {author} {\bibfnamefont {P.}~\bibnamefont
  {Törmä}}\ and\ \bibinfo {author} {\bibfnamefont {W.~L.}\ \bibnamefont
  {Barnes}},\ }\href@noop {} {\bibfield  {journal} {\bibinfo  {journal}
  {Reports on Progress in Physics}\ }\textbf {\bibinfo {volume} {78}},\
  \bibinfo {pages} {013901} (\bibinfo {year} {2014})}\BibitemShut {NoStop}%
\bibitem [{\citenamefont {Bitton}\ \emph {et~al.}(2019)\citenamefont {Bitton},
  \citenamefont {Gupta},\ and\ \citenamefont {Haran}}]{Haranrev2019}%
  \BibitemOpen
  \bibfield  {author} {\bibinfo {author} {\bibfnamefont {O.}~\bibnamefont
  {Bitton}}, \bibinfo {author} {\bibfnamefont {S.~N.}\ \bibnamefont {Gupta}},\
  and\ \bibinfo {author} {\bibfnamefont {G.}~\bibnamefont {Haran}},\
  }\href@noop {} {\bibfield  {journal} {\bibinfo  {journal} {Nanophotonics}\
  }\textbf {\bibinfo {volume} {8}},\ \bibinfo {pages} {559} (\bibinfo {year}
  {2019})}\BibitemShut {NoStop}%
\bibitem [{\citenamefont {Bourgeois}\ \emph {et~al.}(2022)\citenamefont
  {Bourgeois}, \citenamefont {Beutler}, \citenamefont {Khorasani},
  \citenamefont {Panek},\ and\ \citenamefont {Masiello}}]{masiello_2022}%
  \BibitemOpen
  \bibfield  {author} {\bibinfo {author} {\bibfnamefont {M.~R.}\ \bibnamefont
  {Bourgeois}}, \bibinfo {author} {\bibfnamefont {E.~K.}\ \bibnamefont
  {Beutler}}, \bibinfo {author} {\bibfnamefont {S.}~\bibnamefont {Khorasani}},
  \bibinfo {author} {\bibfnamefont {N.}~\bibnamefont {Panek}},\ and\ \bibinfo
  {author} {\bibfnamefont {D.~J.}\ \bibnamefont {Masiello}},\ }\href@noop {}
  {\bibfield  {journal} {\bibinfo  {journal} {Phys. Rev. Lett.}\ }\textbf
  {\bibinfo {volume} {128}},\ \bibinfo {pages} {197401} (\bibinfo {year}
  {2022})}\BibitemShut {NoStop}%
\bibitem [{\citenamefont {Ribeiro}\ \emph {et~al.}(2018)\citenamefont
  {Ribeiro}, \citenamefont {Martínez-Martínez}, \citenamefont {Du},
  \citenamefont {Campos-Gonzalez-Angulo},\ and\ \citenamefont
  {Yuen-Zhou}}]{raphaelrev2018}%
  \BibitemOpen
  \bibfield  {author} {\bibinfo {author} {\bibfnamefont {R.~F.}\ \bibnamefont
  {Ribeiro}}, \bibinfo {author} {\bibfnamefont {L.~A.}\ \bibnamefont
  {Martínez-Martínez}}, \bibinfo {author} {\bibfnamefont {M.}~\bibnamefont
  {Du}}, \bibinfo {author} {\bibfnamefont {J.}~\bibnamefont
  {Campos-Gonzalez-Angulo}},\ and\ \bibinfo {author} {\bibfnamefont
  {J.}~\bibnamefont {Yuen-Zhou}},\ }\href@noop {} {\bibfield  {journal}
  {\bibinfo  {journal} {Chem. Sci.}\ }\textbf {\bibinfo {volume} {9}},\
  \bibinfo {pages} {6325} (\bibinfo {year} {2018})}\BibitemShut {NoStop}%
\bibitem [{\citenamefont {Du}\ \emph {et~al.}(2021)\citenamefont {Du},
  \citenamefont {Campos-Gonzalez-Angulo},\ and\ \citenamefont
  {Yuen-Zhou}}]{mattnoneq2021}%
  \BibitemOpen
  \bibfield  {author} {\bibinfo {author} {\bibfnamefont {M.}~\bibnamefont
  {Du}}, \bibinfo {author} {\bibfnamefont {J.~A.}\ \bibnamefont
  {Campos-Gonzalez-Angulo}},\ and\ \bibinfo {author} {\bibfnamefont
  {J.}~\bibnamefont {Yuen-Zhou}},\ }\href@noop {} {\bibfield  {journal}
  {\bibinfo  {journal} {The Journal of Chemical Physics}\ }\textbf {\bibinfo
  {volume} {154}},\ \bibinfo {pages} {084108} (\bibinfo {year}
  {2021})}\BibitemShut {NoStop}%
\bibitem [{\citenamefont {Dunkelberger}\ \emph
  {et~al.}(2022{\natexlab{a}})\citenamefont {Dunkelberger}, \citenamefont
  {Simpkins}, \citenamefont {Vurgaftman},\ and\ \citenamefont
  {Owrutsky}}]{cina2022}%
  \BibitemOpen
  \bibfield  {author} {\bibinfo {author} {\bibfnamefont {A.~D.}\ \bibnamefont
  {Dunkelberger}}, \bibinfo {author} {\bibfnamefont {B.~S.}\ \bibnamefont
  {Simpkins}}, \bibinfo {author} {\bibfnamefont {I.}~\bibnamefont
  {Vurgaftman}},\ and\ \bibinfo {author} {\bibfnamefont {J.~C.}\ \bibnamefont
  {Owrutsky}},\ }\href@noop {} {\bibfield  {journal} {\bibinfo  {journal}
  {Annual Review of Physical Chemistry}\ }\textbf {\bibinfo {volume} {73}},\
  \bibinfo {pages} {429} (\bibinfo {year} {2022}{\natexlab{a}})}\BibitemShut
  {NoStop}%
\bibitem [{\citenamefont {Hirai}\ \emph
  {et~al.}(2020{\natexlab{a}})\citenamefont {Hirai}, \citenamefont
  {Hutchison},\ and\ \citenamefont {Uji-i}}]{Hiroshirev2020}%
  \BibitemOpen
  \bibfield  {author} {\bibinfo {author} {\bibfnamefont {K.}~\bibnamefont
  {Hirai}}, \bibinfo {author} {\bibfnamefont {J.~A.}\ \bibnamefont
  {Hutchison}},\ and\ \bibinfo {author} {\bibfnamefont {H.}~\bibnamefont
  {Uji-i}},\ }\href@noop {} {\bibfield  {journal} {\bibinfo  {journal}
  {ChemPlusChem}\ }\textbf {\bibinfo {volume} {85}},\ \bibinfo {pages} {1981}
  (\bibinfo {year} {2020}{\natexlab{a}})}\BibitemShut {NoStop}%
\bibitem [{\citenamefont {Du}\ and\ \citenamefont
  {Yuen-Zhou}(2022)}]{mattdark2022}%
  \BibitemOpen
  \bibfield  {author} {\bibinfo {author} {\bibfnamefont {M.}~\bibnamefont
  {Du}}\ and\ \bibinfo {author} {\bibfnamefont {J.}~\bibnamefont {Yuen-Zhou}},\
  }\href@noop {} {\bibfield  {journal} {\bibinfo  {journal} {Phys. Rev. Lett.}\
  }\textbf {\bibinfo {volume} {128}},\ \bibinfo {pages} {096001} (\bibinfo
  {year} {2022})}\BibitemShut {NoStop}%
\bibitem [{\citenamefont {Fraser}(2017)}]{Fraser_2017}%
  \BibitemOpen
  \bibfield  {author} {\bibinfo {author} {\bibfnamefont {M.~D.}\ \bibnamefont
  {Fraser}},\ }\href@noop {} {\bibfield  {journal} {\bibinfo  {journal}
  {Semiconductor Science and Technology}\ }\textbf {\bibinfo {volume} {32}},\
  \bibinfo {pages} {093003} (\bibinfo {year} {2017})}\BibitemShut {NoStop}%
\bibitem [{\citenamefont {Berini}\ and\ \citenamefont
  {De~Leon}(2012)}]{beriniamp2012}%
  \BibitemOpen
  \bibfield  {author} {\bibinfo {author} {\bibfnamefont {P.}~\bibnamefont
  {Berini}}\ and\ \bibinfo {author} {\bibfnamefont {I.}~\bibnamefont
  {De~Leon}},\ }\href@noop {} {\bibfield  {journal} {\bibinfo  {journal}
  {Nature Photonics}\ }\textbf {\bibinfo {volume} {6}},\ \bibinfo {pages} {16}
  (\bibinfo {year} {2012})}\BibitemShut {NoStop}%
\bibitem [{\citenamefont {Jamadi}\ \emph {et~al.}(2018)\citenamefont {Jamadi},
  \citenamefont {Reveret}, \citenamefont {Disseix}, \citenamefont {Medard},
  \citenamefont {Leymarie}, \citenamefont {Moreau}, \citenamefont {Solnyshkov},
  \citenamefont {Deparis}, \citenamefont {Leroux}, \citenamefont {Cambril},
  \citenamefont {Bouchoule}, \citenamefont {Zuniga-Perez},\ and\ \citenamefont
  {Malpuech}}]{Jamadiamp2018}%
  \BibitemOpen
  \bibfield  {author} {\bibinfo {author} {\bibfnamefont {O.}~\bibnamefont
  {Jamadi}}, \bibinfo {author} {\bibfnamefont {F.}~\bibnamefont {Reveret}},
  \bibinfo {author} {\bibfnamefont {P.}~\bibnamefont {Disseix}}, \bibinfo
  {author} {\bibfnamefont {F.}~\bibnamefont {Medard}}, \bibinfo {author}
  {\bibfnamefont {J.}~\bibnamefont {Leymarie}}, \bibinfo {author}
  {\bibfnamefont {A.}~\bibnamefont {Moreau}}, \bibinfo {author} {\bibfnamefont
  {D.}~\bibnamefont {Solnyshkov}}, \bibinfo {author} {\bibfnamefont
  {C.}~\bibnamefont {Deparis}}, \bibinfo {author} {\bibfnamefont
  {M.}~\bibnamefont {Leroux}}, \bibinfo {author} {\bibfnamefont
  {E.}~\bibnamefont {Cambril}}, \bibinfo {author} {\bibfnamefont
  {S.}~\bibnamefont {Bouchoule}}, \bibinfo {author} {\bibfnamefont
  {J.}~\bibnamefont {Zuniga-Perez}},\ and\ \bibinfo {author} {\bibfnamefont
  {G.}~\bibnamefont {Malpuech}},\ }\href@noop {} {\bibfield  {journal}
  {\bibinfo  {journal} {Light: Science {\&} Applications}\ }\textbf {\bibinfo
  {volume} {7}},\ \bibinfo {pages} {82} (\bibinfo {year} {2018})}\BibitemShut
  {NoStop}%
\bibitem [{\citenamefont {Nguyen}\ \emph {et~al.}(2013)\citenamefont {Nguyen},
  \citenamefont {Vishnevsky}, \citenamefont {Sturm}, \citenamefont {Tanese},
  \citenamefont {Solnyshkov}, \citenamefont {Galopin}, \citenamefont
  {Lema\^{\i}tre}, \citenamefont {Sagnes}, \citenamefont {Amo}, \citenamefont
  {Malpuech},\ and\ \citenamefont {Bloch}}]{Blochdiode2013}%
  \BibitemOpen
  \bibfield  {author} {\bibinfo {author} {\bibfnamefont {H.~S.}\ \bibnamefont
  {Nguyen}}, \bibinfo {author} {\bibfnamefont {D.}~\bibnamefont {Vishnevsky}},
  \bibinfo {author} {\bibfnamefont {C.}~\bibnamefont {Sturm}}, \bibinfo
  {author} {\bibfnamefont {D.}~\bibnamefont {Tanese}}, \bibinfo {author}
  {\bibfnamefont {D.}~\bibnamefont {Solnyshkov}}, \bibinfo {author}
  {\bibfnamefont {E.}~\bibnamefont {Galopin}}, \bibinfo {author} {\bibfnamefont
  {A.}~\bibnamefont {Lema\^{\i}tre}}, \bibinfo {author} {\bibfnamefont
  {I.}~\bibnamefont {Sagnes}}, \bibinfo {author} {\bibfnamefont
  {A.}~\bibnamefont {Amo}}, \bibinfo {author} {\bibfnamefont {G.}~\bibnamefont
  {Malpuech}},\ and\ \bibinfo {author} {\bibfnamefont {J.}~\bibnamefont
  {Bloch}},\ }\href@noop {} {\bibfield  {journal} {\bibinfo  {journal} {Phys.
  Rev. Lett.}\ }\textbf {\bibinfo {volume} {110}},\ \bibinfo {pages} {236601}
  (\bibinfo {year} {2013})}\BibitemShut {NoStop}%
\bibitem [{\citenamefont {Marsault}\ \emph {et~al.}(2015)\citenamefont
  {Marsault}, \citenamefont {Nguyen}, \citenamefont {Tanese}, \citenamefont
  {Lemaître}, \citenamefont {Galopin}, \citenamefont {Sagnes}, \citenamefont
  {Amo},\ and\ \citenamefont {Bloch}}]{Felix_rou_2015}%
  \BibitemOpen
  \bibfield  {author} {\bibinfo {author} {\bibfnamefont {F.}~\bibnamefont
  {Marsault}}, \bibinfo {author} {\bibfnamefont {H.~S.}\ \bibnamefont
  {Nguyen}}, \bibinfo {author} {\bibfnamefont {D.}~\bibnamefont {Tanese}},
  \bibinfo {author} {\bibfnamefont {A.}~\bibnamefont {Lemaître}}, \bibinfo
  {author} {\bibfnamefont {E.}~\bibnamefont {Galopin}}, \bibinfo {author}
  {\bibfnamefont {I.}~\bibnamefont {Sagnes}}, \bibinfo {author} {\bibfnamefont
  {A.}~\bibnamefont {Amo}},\ and\ \bibinfo {author} {\bibfnamefont
  {J.}~\bibnamefont {Bloch}},\ }\href@noop {} {\bibfield  {journal} {\bibinfo
  {journal} {Applied Physics Letters}\ }\textbf {\bibinfo {volume} {107}},\
  \bibinfo {pages} {201115} (\bibinfo {year} {2015})}\BibitemShut {NoStop}%
\bibitem [{\citenamefont {Amo}\ \emph {et~al.}(2010)\citenamefont {Amo},
  \citenamefont {Liew}, \citenamefont {Adrados}, \citenamefont {Houdr{\'e}},
  \citenamefont {Giacobino}, \citenamefont {Kavokin},\ and\ \citenamefont
  {Bramati}}]{Amo2010}%
  \BibitemOpen
  \bibfield  {author} {\bibinfo {author} {\bibfnamefont {A.}~\bibnamefont
  {Amo}}, \bibinfo {author} {\bibfnamefont {T.~C.~H.}\ \bibnamefont {Liew}},
  \bibinfo {author} {\bibfnamefont {C.}~\bibnamefont {Adrados}}, \bibinfo
  {author} {\bibfnamefont {R.}~\bibnamefont {Houdr{\'e}}}, \bibinfo {author}
  {\bibfnamefont {E.}~\bibnamefont {Giacobino}}, \bibinfo {author}
  {\bibfnamefont {A.~V.}\ \bibnamefont {Kavokin}},\ and\ \bibinfo {author}
  {\bibfnamefont {A.}~\bibnamefont {Bramati}},\ }\href@noop {} {\bibfield
  {journal} {\bibinfo  {journal} {Nature Photonics}\ }\textbf {\bibinfo
  {volume} {4}},\ \bibinfo {pages} {361} (\bibinfo {year} {2010})}\BibitemShut
  {NoStop}%
\bibitem [{\citenamefont {Chen}\ \emph {et~al.}(2022)\citenamefont {Chen},
  \citenamefont {Li}, \citenamefont {Zhou}, \citenamefont {Luo}, \citenamefont
  {Sun}, \citenamefont {Ye}, \citenamefont {Sun}, \citenamefont {Wang},
  \citenamefont {Zheng}, \citenamefont {Chen}, \citenamefont {Xu},
  \citenamefont {Xu}, \citenamefont {Byrnes}, \citenamefont {Chen},\ and\
  \citenamefont {Wu}}]{Chen_sw_2022}%
  \BibitemOpen
  \bibfield  {author} {\bibinfo {author} {\bibfnamefont {F.}~\bibnamefont
  {Chen}}, \bibinfo {author} {\bibfnamefont {H.}~\bibnamefont {Li}}, \bibinfo
  {author} {\bibfnamefont {H.}~\bibnamefont {Zhou}}, \bibinfo {author}
  {\bibfnamefont {S.}~\bibnamefont {Luo}}, \bibinfo {author} {\bibfnamefont
  {Z.}~\bibnamefont {Sun}}, \bibinfo {author} {\bibfnamefont {Z.}~\bibnamefont
  {Ye}}, \bibinfo {author} {\bibfnamefont {F.}~\bibnamefont {Sun}}, \bibinfo
  {author} {\bibfnamefont {J.}~\bibnamefont {Wang}}, \bibinfo {author}
  {\bibfnamefont {Y.}~\bibnamefont {Zheng}}, \bibinfo {author} {\bibfnamefont
  {X.}~\bibnamefont {Chen}}, \bibinfo {author} {\bibfnamefont {H.}~\bibnamefont
  {Xu}}, \bibinfo {author} {\bibfnamefont {H.}~\bibnamefont {Xu}}, \bibinfo
  {author} {\bibfnamefont {T.}~\bibnamefont {Byrnes}}, \bibinfo {author}
  {\bibfnamefont {Z.}~\bibnamefont {Chen}},\ and\ \bibinfo {author}
  {\bibfnamefont {J.}~\bibnamefont {Wu}},\ }\href@noop {} {\bibfield  {journal}
  {\bibinfo  {journal} {Phys. Rev. Lett.}\ }\textbf {\bibinfo {volume} {129}},\
  \bibinfo {pages} {057402} (\bibinfo {year} {2022})}\BibitemShut {NoStop}%
\bibitem [{\citenamefont {Pscherer}\ \emph {et~al.}(2021)\citenamefont
  {Pscherer}, \citenamefont {Meierhofer}, \citenamefont {Wang}, \citenamefont
  {Kelkar}, \citenamefont {Mart\'{\i}n-Cano}, \citenamefont {Utikal},
  \citenamefont {G\"otzinger},\ and\ \citenamefont {Sandoghdar}}]{vahid_2021}%
  \BibitemOpen
  \bibfield  {author} {\bibinfo {author} {\bibfnamefont {A.}~\bibnamefont
  {Pscherer}}, \bibinfo {author} {\bibfnamefont {M.}~\bibnamefont
  {Meierhofer}}, \bibinfo {author} {\bibfnamefont {D.}~\bibnamefont {Wang}},
  \bibinfo {author} {\bibfnamefont {H.}~\bibnamefont {Kelkar}}, \bibinfo
  {author} {\bibfnamefont {D.}~\bibnamefont {Mart\'{\i}n-Cano}}, \bibinfo
  {author} {\bibfnamefont {T.}~\bibnamefont {Utikal}}, \bibinfo {author}
  {\bibfnamefont {S.}~\bibnamefont {G\"otzinger}},\ and\ \bibinfo {author}
  {\bibfnamefont {V.}~\bibnamefont {Sandoghdar}},\ }\href@noop {} {\bibfield
  {journal} {\bibinfo  {journal} {Phys. Rev. Lett.}\ }\textbf {\bibinfo
  {volume} {127}},\ \bibinfo {pages} {133603} (\bibinfo {year}
  {2021})}\BibitemShut {NoStop}%
\bibitem [{\citenamefont {Wang}\ \emph {et~al.}(2021)\citenamefont {Wang},
  \citenamefont {Hertzog},\ and\ \citenamefont {B{\"o}rjesson}}]{Wang2021}%
  \BibitemOpen
  \bibfield  {author} {\bibinfo {author} {\bibfnamefont {M.}~\bibnamefont
  {Wang}}, \bibinfo {author} {\bibfnamefont {M.}~\bibnamefont {Hertzog}},\ and\
  \bibinfo {author} {\bibfnamefont {K.}~\bibnamefont {B{\"o}rjesson}},\
  }\href@noop {} {\bibfield  {journal} {\bibinfo  {journal} {Nature
  Communications}\ }\textbf {\bibinfo {volume} {12}},\ \bibinfo {pages} {1874}
  (\bibinfo {year} {2021})}\BibitemShut {NoStop}%
\bibitem [{\citenamefont {Du}\ \emph {et~al.}(2018)\citenamefont {Du},
  \citenamefont {Martínez-Martínez}, \citenamefont {Ribeiro}, \citenamefont
  {Hu}, \citenamefont {Menon},\ and\ \citenamefont
  {Yuen-Zhou}}]{mattparet_2018}%
  \BibitemOpen
  \bibfield  {author} {\bibinfo {author} {\bibfnamefont {M.}~\bibnamefont
  {Du}}, \bibinfo {author} {\bibfnamefont {L.~A.}\ \bibnamefont
  {Martínez-Martínez}}, \bibinfo {author} {\bibfnamefont {R.~F.}\
  \bibnamefont {Ribeiro}}, \bibinfo {author} {\bibfnamefont {Z.}~\bibnamefont
  {Hu}}, \bibinfo {author} {\bibfnamefont {V.~M.}\ \bibnamefont {Menon}},\ and\
  \bibinfo {author} {\bibfnamefont {J.}~\bibnamefont {Yuen-Zhou}},\ }\href@noop
  {} {\bibfield  {journal} {\bibinfo  {journal} {Chem. Sci.}\ }\textbf
  {\bibinfo {volume} {9}},\ \bibinfo {pages} {6659} (\bibinfo {year}
  {2018})}\BibitemShut {NoStop}%
\bibitem [{\citenamefont {Groenhof}\ and\ \citenamefont
  {Toppari}(2018)}]{groenhoff_2018}%
  \BibitemOpen
  \bibfield  {author} {\bibinfo {author} {\bibfnamefont {G.}~\bibnamefont
  {Groenhof}}\ and\ \bibinfo {author} {\bibfnamefont {J.~J.}\ \bibnamefont
  {Toppari}},\ }\href@noop {} {\bibfield  {journal} {\bibinfo  {journal} {The
  Journal of Physical Chemistry Letters}\ }\textbf {\bibinfo {volume} {9}},\
  \bibinfo {pages} {4848} (\bibinfo {year} {2018})}\BibitemShut {NoStop}%
\bibitem [{\citenamefont {Thomas}\ \emph {et~al.}(2019)\citenamefont {Thomas},
  \citenamefont {Lethuillier-Karl}, \citenamefont {Nagarajan}, \citenamefont
  {Vergauwe}, \citenamefont {George}, \citenamefont {Chervy}, \citenamefont
  {Shalabney}, \citenamefont {Devaux}, \citenamefont {Genet}, \citenamefont
  {Moran},\ and\ \citenamefont {Ebbesen}}]{Thomas2019}%
  \BibitemOpen
  \bibfield  {author} {\bibinfo {author} {\bibfnamefont {A.}~\bibnamefont
  {Thomas}}, \bibinfo {author} {\bibfnamefont {L.}~\bibnamefont
  {Lethuillier-Karl}}, \bibinfo {author} {\bibfnamefont {K.}~\bibnamefont
  {Nagarajan}}, \bibinfo {author} {\bibfnamefont {R.~M.~A.}\ \bibnamefont
  {Vergauwe}}, \bibinfo {author} {\bibfnamefont {J.}~\bibnamefont {George}},
  \bibinfo {author} {\bibfnamefont {T.}~\bibnamefont {Chervy}}, \bibinfo
  {author} {\bibfnamefont {A.}~\bibnamefont {Shalabney}}, \bibinfo {author}
  {\bibfnamefont {E.}~\bibnamefont {Devaux}}, \bibinfo {author} {\bibfnamefont
  {C.}~\bibnamefont {Genet}}, \bibinfo {author} {\bibfnamefont
  {J.}~\bibnamefont {Moran}},\ and\ \bibinfo {author} {\bibfnamefont {T.~W.}\
  \bibnamefont {Ebbesen}},\ }\href@noop {} {\bibfield  {journal} {\bibinfo
  {journal} {Science}\ }\textbf {\bibinfo {volume} {363}},\ \bibinfo {pages}
  {615} (\bibinfo {year} {2019})}\BibitemShut {NoStop}%
\bibitem [{\citenamefont {Hirai}\ \emph
  {et~al.}(2020{\natexlab{b}})\citenamefont {Hirai}, \citenamefont {Takeda},
  \citenamefont {Hutchison},\ and\ \citenamefont {Uji-i}}]{Hirai2020}%
  \BibitemOpen
  \bibfield  {author} {\bibinfo {author} {\bibfnamefont {K.}~\bibnamefont
  {Hirai}}, \bibinfo {author} {\bibfnamefont {R.}~\bibnamefont {Takeda}},
  \bibinfo {author} {\bibfnamefont {J.~A.}\ \bibnamefont {Hutchison}},\ and\
  \bibinfo {author} {\bibfnamefont {H.}~\bibnamefont {Uji-i}},\ }\href@noop {}
  {\bibfield  {journal} {\bibinfo  {journal} {Angewandte Chemie International
  Edition}\ }\textbf {\bibinfo {volume} {59}},\ \bibinfo {pages} {5332}
  (\bibinfo {year} {2020}{\natexlab{b}})}\BibitemShut {NoStop}%
\bibitem [{\citenamefont {Du}\ \emph {et~al.}(2019)\citenamefont {Du},
  \citenamefont {Ribero},\ and\ \citenamefont {Yuen-Zhou}}]{matt_remote_2019}%
  \BibitemOpen
  \bibfield  {author} {\bibinfo {author} {\bibfnamefont {M.}~\bibnamefont
  {Du}}, \bibinfo {author} {\bibfnamefont {R.~F.}\ \bibnamefont {Ribero}},\
  and\ \bibinfo {author} {\bibfnamefont {J.}~\bibnamefont {Yuen-Zhou}},\
  }\href@noop {} {\bibfield  {journal} {\bibinfo  {journal} {Chem}\ }\textbf
  {\bibinfo {volume} {5}},\ \bibinfo {pages} {1167} (\bibinfo {year}
  {2019})}\BibitemShut {NoStop}%
\bibitem [{\citenamefont {Haugland}\ \emph {et~al.}(2020)\citenamefont
  {Haugland}, \citenamefont {Ronca}, \citenamefont {Kj\o{}nstad}, \citenamefont
  {Rubio},\ and\ \citenamefont {Koch}}]{Rubio_2020}%
  \BibitemOpen
  \bibfield  {author} {\bibinfo {author} {\bibfnamefont {T.~S.}\ \bibnamefont
  {Haugland}}, \bibinfo {author} {\bibfnamefont {E.}~\bibnamefont {Ronca}},
  \bibinfo {author} {\bibfnamefont {E.~F.}\ \bibnamefont {Kj\o{}nstad}},
  \bibinfo {author} {\bibfnamefont {A.}~\bibnamefont {Rubio}},\ and\ \bibinfo
  {author} {\bibfnamefont {H.}~\bibnamefont {Koch}},\ }\href@noop {} {\bibfield
   {journal} {\bibinfo  {journal} {Phys. Rev. X}\ }\textbf {\bibinfo {volume}
  {10}},\ \bibinfo {pages} {041043} (\bibinfo {year} {2020})}\BibitemShut
  {NoStop}%
\bibitem [{\citenamefont {Sidler}\ \emph {et~al.}(2021)\citenamefont {Sidler},
  \citenamefont {Schäfer}, \citenamefont {Ruggenthaler},\ and\ \citenamefont
  {Rubio}}]{Rubio1_2021}%
  \BibitemOpen
  \bibfield  {author} {\bibinfo {author} {\bibfnamefont {D.}~\bibnamefont
  {Sidler}}, \bibinfo {author} {\bibfnamefont {C.}~\bibnamefont {Schäfer}},
  \bibinfo {author} {\bibfnamefont {M.}~\bibnamefont {Ruggenthaler}},\ and\
  \bibinfo {author} {\bibfnamefont {A.}~\bibnamefont {Rubio}},\ }\href@noop {}
  {\bibfield  {journal} {\bibinfo  {journal} {The Journal of Physical Chemistry
  Letters}\ }\textbf {\bibinfo {volume} {12}},\ \bibinfo {pages} {508}
  (\bibinfo {year} {2021})}\BibitemShut {NoStop}%
\bibitem [{\citenamefont {Basov}\ \emph {et~al.}(2021)\citenamefont {Basov},
  \citenamefont {Asenjo-Garcia}, \citenamefont {Schuck}, \citenamefont {Zhu},\
  and\ \citenamefont {Rubio}}]{Rubio2_2021}%
  \BibitemOpen
  \bibfield  {author} {\bibinfo {author} {\bibfnamefont {D.~N.}\ \bibnamefont
  {Basov}}, \bibinfo {author} {\bibfnamefont {A.}~\bibnamefont
  {Asenjo-Garcia}}, \bibinfo {author} {\bibfnamefont {P.~J.}\ \bibnamefont
  {Schuck}}, \bibinfo {author} {\bibfnamefont {X.}~\bibnamefont {Zhu}},\ and\
  \bibinfo {author} {\bibfnamefont {A.}~\bibnamefont {Rubio}},\ }\href@noop {}
  {\bibfield  {journal} {\bibinfo  {journal} {Nanophotonics}\ }\textbf
  {\bibinfo {volume} {10}},\ \bibinfo {pages} {549} (\bibinfo {year}
  {2021})}\BibitemShut {NoStop}%
\bibitem [{\citenamefont {May}\ \emph {et~al.}(2020)\citenamefont {May},
  \citenamefont {Fialkow}, \citenamefont {Wu}, \citenamefont {Park},
  \citenamefont {Leng}, \citenamefont {Kropp}, \citenamefont {Gougousi},
  \citenamefont {Lalanne}, \citenamefont {Pelton},\ and\ \citenamefont
  {Raschke}}]{pelton_2020}%
  \BibitemOpen
  \bibfield  {author} {\bibinfo {author} {\bibfnamefont {M.~A.}\ \bibnamefont
  {May}}, \bibinfo {author} {\bibfnamefont {D.}~\bibnamefont {Fialkow}},
  \bibinfo {author} {\bibfnamefont {T.}~\bibnamefont {Wu}}, \bibinfo {author}
  {\bibfnamefont {K.-D.}\ \bibnamefont {Park}}, \bibinfo {author}
  {\bibfnamefont {H.}~\bibnamefont {Leng}}, \bibinfo {author} {\bibfnamefont
  {J.~A.}\ \bibnamefont {Kropp}}, \bibinfo {author} {\bibfnamefont
  {T.}~\bibnamefont {Gougousi}}, \bibinfo {author} {\bibfnamefont
  {P.}~\bibnamefont {Lalanne}}, \bibinfo {author} {\bibfnamefont
  {M.}~\bibnamefont {Pelton}},\ and\ \bibinfo {author} {\bibfnamefont {M.~B.}\
  \bibnamefont {Raschke}},\ }\href@noop {} {\bibfield  {journal} {\bibinfo
  {journal} {Advanced Quantum Technologies}\ }\textbf {\bibinfo {volume} {3}},\
  \bibinfo {pages} {1900087} (\bibinfo {year} {2020})}\BibitemShut {NoStop}%
\bibitem [{\citenamefont {Li}\ \emph {et~al.}(2022)\citenamefont {Li},
  \citenamefont {Cui}, \citenamefont {Subotnik},\ and\ \citenamefont
  {Nitzan}}]{taoli_2022}%
  \BibitemOpen
  \bibfield  {author} {\bibinfo {author} {\bibfnamefont {T.~E.}\ \bibnamefont
  {Li}}, \bibinfo {author} {\bibfnamefont {B.}~\bibnamefont {Cui}}, \bibinfo
  {author} {\bibfnamefont {J.~E.}\ \bibnamefont {Subotnik}},\ and\ \bibinfo
  {author} {\bibfnamefont {A.}~\bibnamefont {Nitzan}},\ }\href@noop {}
  {\bibfield  {journal} {\bibinfo  {journal} {Annual Review of Physical
  Chemistry}\ }\textbf {\bibinfo {volume} {73}},\ \bibinfo {pages} {43}
  (\bibinfo {year} {2022})}\BibitemShut {NoStop}%
\bibitem [{\citenamefont {Keeling}\ and\ \citenamefont
  {K\'{e}na-Cohen}(2020)}]{keeling_rev_2020}%
  \BibitemOpen
  \bibfield  {author} {\bibinfo {author} {\bibfnamefont {J.}~\bibnamefont
  {Keeling}}\ and\ \bibinfo {author} {\bibfnamefont {S.}~\bibnamefont
  {K\'{e}na-Cohen}},\ }\href@noop {} {\bibfield  {journal} {\bibinfo  {journal}
  {Annual Review of Physical Chemistry}\ }\textbf {\bibinfo {volume} {71}},\
  \bibinfo {pages} {435} (\bibinfo {year} {2020})}\BibitemShut {NoStop}%
\bibitem [{\citenamefont {Cortese}\ \emph {et~al.}(2017)\citenamefont
  {Cortese}, \citenamefont {Lagoudakis},\ and\ \citenamefont
  {De~Liberato}}]{deliberato_2017}%
  \BibitemOpen
  \bibfield  {author} {\bibinfo {author} {\bibfnamefont {E.}~\bibnamefont
  {Cortese}}, \bibinfo {author} {\bibfnamefont {P.~G.}\ \bibnamefont
  {Lagoudakis}},\ and\ \bibinfo {author} {\bibfnamefont {S.}~\bibnamefont
  {De~Liberato}},\ }\href@noop {} {\bibfield  {journal} {\bibinfo  {journal}
  {Phys. Rev. Lett.}\ }\textbf {\bibinfo {volume} {119}},\ \bibinfo {pages}
  {043604} (\bibinfo {year} {2017})}\BibitemShut {NoStop}%
\bibitem [{\citenamefont {Chikkaraddy}\ \emph {et~al.}(2016)\citenamefont
  {Chikkaraddy}, \citenamefont {de~Nijs}, \citenamefont {Benz}, \citenamefont
  {Barrow}, \citenamefont {Scherman}, \citenamefont {Rosta}, \citenamefont
  {Demetriadou}, \citenamefont {Fox}, \citenamefont {Hess},\ and\ \citenamefont
  {Baumberg}}]{Chikkaraddy2016}%
  \BibitemOpen
  \bibfield  {author} {\bibinfo {author} {\bibfnamefont {R.}~\bibnamefont
  {Chikkaraddy}}, \bibinfo {author} {\bibfnamefont {B.}~\bibnamefont
  {de~Nijs}}, \bibinfo {author} {\bibfnamefont {F.}~\bibnamefont {Benz}},
  \bibinfo {author} {\bibfnamefont {S.~J.}\ \bibnamefont {Barrow}}, \bibinfo
  {author} {\bibfnamefont {O.~A.}\ \bibnamefont {Scherman}}, \bibinfo {author}
  {\bibfnamefont {E.}~\bibnamefont {Rosta}}, \bibinfo {author} {\bibfnamefont
  {A.}~\bibnamefont {Demetriadou}}, \bibinfo {author} {\bibfnamefont
  {P.}~\bibnamefont {Fox}}, \bibinfo {author} {\bibfnamefont {O.}~\bibnamefont
  {Hess}},\ and\ \bibinfo {author} {\bibfnamefont {J.~J.}\ \bibnamefont
  {Baumberg}},\ }\href@noop {} {\bibfield  {journal} {\bibinfo  {journal}
  {Nature}\ }\textbf {\bibinfo {volume} {535}},\ \bibinfo {pages} {127}
  (\bibinfo {year} {2016})}\BibitemShut {NoStop}%
\bibitem [{\citenamefont {Zasedatelev}\ \emph {et~al.}(2019)\citenamefont
  {Zasedatelev}, \citenamefont {Baranikov}, \citenamefont {Urbonas},
  \citenamefont {Scafirimuto}, \citenamefont {Scherf}, \citenamefont
  {St{\"o}ferle}, \citenamefont {Mahrt},\ and\ \citenamefont
  {Lagoudakis}}]{laguodakis2019}%
  \BibitemOpen
  \bibfield  {author} {\bibinfo {author} {\bibfnamefont {A.~V.}\ \bibnamefont
  {Zasedatelev}}, \bibinfo {author} {\bibfnamefont {A.~V.}\ \bibnamefont
  {Baranikov}}, \bibinfo {author} {\bibfnamefont {D.}~\bibnamefont {Urbonas}},
  \bibinfo {author} {\bibfnamefont {F.}~\bibnamefont {Scafirimuto}}, \bibinfo
  {author} {\bibfnamefont {U.}~\bibnamefont {Scherf}}, \bibinfo {author}
  {\bibfnamefont {T.}~\bibnamefont {St{\"o}ferle}}, \bibinfo {author}
  {\bibfnamefont {R.~F.}\ \bibnamefont {Mahrt}},\ and\ \bibinfo {author}
  {\bibfnamefont {P.~G.}\ \bibnamefont {Lagoudakis}},\ }\href@noop {}
  {\bibfield  {journal} {\bibinfo  {journal} {Nature Photonics}\ }\textbf
  {\bibinfo {volume} {13}},\ \bibinfo {pages} {378} (\bibinfo {year}
  {2019})}\BibitemShut {NoStop}%
\bibitem [{\citenamefont {K{\'e}na-Cohen}\ and\ \citenamefont
  {Forrest}(2010)}]{Kena-Cohen2010}%
  \BibitemOpen
  \bibfield  {author} {\bibinfo {author} {\bibfnamefont {S.}~\bibnamefont
  {K{\'e}na-Cohen}}\ and\ \bibinfo {author} {\bibfnamefont {S.~R.}\
  \bibnamefont {Forrest}},\ }\href@noop {} {\bibfield  {journal} {\bibinfo
  {journal} {Nature Photonics}\ }\textbf {\bibinfo {volume} {4}},\ \bibinfo
  {pages} {371} (\bibinfo {year} {2010})}\BibitemShut {NoStop}%
\bibitem [{\citenamefont {Baumberg}\ \emph {et~al.}(2008)\citenamefont
  {Baumberg}, \citenamefont {Kavokin}, \citenamefont {Christopoulos},
  \citenamefont {Grundy}, \citenamefont {Butt\'e}, \citenamefont {Christmann},
  \citenamefont {Solnyshkov}, \citenamefont {Malpuech}, \citenamefont
  {Baldassarri H\"oger~von H\"ogersthal}, \citenamefont {Feltin}, \citenamefont
  {Carlin},\ and\ \citenamefont {Grandjean}}]{Baumberg2008}%
  \BibitemOpen
  \bibfield  {author} {\bibinfo {author} {\bibfnamefont {J.~J.}\ \bibnamefont
  {Baumberg}}, \bibinfo {author} {\bibfnamefont {A.~V.}\ \bibnamefont
  {Kavokin}}, \bibinfo {author} {\bibfnamefont {S.}~\bibnamefont
  {Christopoulos}}, \bibinfo {author} {\bibfnamefont {A.~J.~D.}\ \bibnamefont
  {Grundy}}, \bibinfo {author} {\bibfnamefont {R.}~\bibnamefont {Butt\'e}},
  \bibinfo {author} {\bibfnamefont {G.}~\bibnamefont {Christmann}}, \bibinfo
  {author} {\bibfnamefont {D.~D.}\ \bibnamefont {Solnyshkov}}, \bibinfo
  {author} {\bibfnamefont {G.}~\bibnamefont {Malpuech}}, \bibinfo {author}
  {\bibfnamefont {G.}~\bibnamefont {Baldassarri H\"oger~von H\"ogersthal}},
  \bibinfo {author} {\bibfnamefont {E.}~\bibnamefont {Feltin}}, \bibinfo
  {author} {\bibfnamefont {J.-F.}\ \bibnamefont {Carlin}},\ and\ \bibinfo
  {author} {\bibfnamefont {N.}~\bibnamefont {Grandjean}},\ }\href@noop {}
  {\bibfield  {journal} {\bibinfo  {journal} {Phys. Rev. Lett.}\ }\textbf
  {\bibinfo {volume} {101}},\ \bibinfo {pages} {136409} (\bibinfo {year}
  {2008})}\BibitemShut {NoStop}%
\bibitem [{\citenamefont {Ghosh}\ and\ \citenamefont {Liew}(2020)}]{Ghosh2020}%
  \BibitemOpen
  \bibfield  {author} {\bibinfo {author} {\bibfnamefont {S.}~\bibnamefont
  {Ghosh}}\ and\ \bibinfo {author} {\bibfnamefont {T.~C.~H.}\ \bibnamefont
  {Liew}},\ }\href@noop {} {\bibfield  {journal} {\bibinfo  {journal} {npj
  Quantum Information}\ }\textbf {\bibinfo {volume} {6}},\ \bibinfo {pages}
  {16} (\bibinfo {year} {2020})}\BibitemShut {NoStop}%
\bibitem [{\citenamefont {Pelton}\ \emph {et~al.}(2019)\citenamefont {Pelton},
  \citenamefont {Storm},\ and\ \citenamefont {Leng}}]{pelton_2019}%
  \BibitemOpen
  \bibfield  {author} {\bibinfo {author} {\bibfnamefont {M.}~\bibnamefont
  {Pelton}}, \bibinfo {author} {\bibfnamefont {S.~D.}\ \bibnamefont {Storm}},\
  and\ \bibinfo {author} {\bibfnamefont {H.}~\bibnamefont {Leng}},\ }\href@noop
  {} {\bibfield  {journal} {\bibinfo  {journal} {Nanoscale}\ }\textbf {\bibinfo
  {volume} {11}},\ \bibinfo {pages} {14540} (\bibinfo {year}
  {2019})}\BibitemShut {NoStop}%
\bibitem [{\citenamefont {Pannir-Sivajothi}\ \emph {et~al.}(2022)\citenamefont
  {Pannir-Sivajothi}, \citenamefont {Campos-Gonzalez-Angulo}, \citenamefont
  {Mart{\'i}nez-Mart{\'i}nez}, \citenamefont {Sinha},\ and\ \citenamefont
  {Yuen-Zhou}}]{sindhana2022}%
  \BibitemOpen
  \bibfield  {author} {\bibinfo {author} {\bibfnamefont {S.}~\bibnamefont
  {Pannir-Sivajothi}}, \bibinfo {author} {\bibfnamefont {J.~A.}\ \bibnamefont
  {Campos-Gonzalez-Angulo}}, \bibinfo {author} {\bibfnamefont {L.~A.}\
  \bibnamefont {Mart{\'i}nez-Mart{\'i}nez}}, \bibinfo {author} {\bibfnamefont
  {S.}~\bibnamefont {Sinha}},\ and\ \bibinfo {author} {\bibfnamefont
  {J.}~\bibnamefont {Yuen-Zhou}},\ }\href@noop {} {\bibfield  {journal}
  {\bibinfo  {journal} {Nature Communications}\ }\textbf {\bibinfo {volume}
  {13}},\ \bibinfo {pages} {1645} (\bibinfo {year} {2022})}\BibitemShut
  {NoStop}%
\bibitem [{\citenamefont {Mondal}\ \emph {et~al.}(2022)\citenamefont {Mondal},
  \citenamefont {Semenov}, \citenamefont {Ochoa},\ and\ \citenamefont
  {Nitzan}}]{Nitzan_ir_2022}%
  \BibitemOpen
  \bibfield  {author} {\bibinfo {author} {\bibfnamefont {M.}~\bibnamefont
  {Mondal}}, \bibinfo {author} {\bibfnamefont {A.}~\bibnamefont {Semenov}},
  \bibinfo {author} {\bibfnamefont {M.~A.}\ \bibnamefont {Ochoa}},\ and\
  \bibinfo {author} {\bibfnamefont {A.}~\bibnamefont {Nitzan}},\ }\href@noop {}
  {\bibfield  {journal} {\bibinfo  {journal} {The Journal of Physical Chemistry
  Letters}\ }\textbf {\bibinfo {volume} {13}},\ \bibinfo {pages} {9673}
  (\bibinfo {year} {2022})}\BibitemShut {NoStop}%
\bibitem [{\citenamefont {Aspelmeyer}\ \emph {et~al.}(2014)\citenamefont
  {Aspelmeyer}, \citenamefont {Kippenberg},\ and\ \citenamefont
  {Marquardt}}]{opt_mech_rev_2014}%
  \BibitemOpen
  \bibfield  {author} {\bibinfo {author} {\bibfnamefont {M.}~\bibnamefont
  {Aspelmeyer}}, \bibinfo {author} {\bibfnamefont {T.~J.}\ \bibnamefont
  {Kippenberg}},\ and\ \bibinfo {author} {\bibfnamefont {F.}~\bibnamefont
  {Marquardt}},\ }\href@noop {} {\bibfield  {journal} {\bibinfo  {journal}
  {Rev. Mod. Phys.}\ }\textbf {\bibinfo {volume} {86}},\ \bibinfo {pages}
  {1391} (\bibinfo {year} {2014})}\BibitemShut {NoStop}%
\bibitem [{\citenamefont {Elste}\ \emph {et~al.}(2009)\citenamefont {Elste},
  \citenamefont {Girvin},\ and\ \citenamefont {Clerk}}]{Clerk1_2019}%
  \BibitemOpen
  \bibfield  {author} {\bibinfo {author} {\bibfnamefont {F.}~\bibnamefont
  {Elste}}, \bibinfo {author} {\bibfnamefont {S.~M.}\ \bibnamefont {Girvin}},\
  and\ \bibinfo {author} {\bibfnamefont {A.~A.}\ \bibnamefont {Clerk}},\
  }\href@noop {} {\bibfield  {journal} {\bibinfo  {journal} {Phys. Rev. Lett.}\
  }\textbf {\bibinfo {volume} {102}},\ \bibinfo {pages} {207209} (\bibinfo
  {year} {2009})}\BibitemShut {NoStop}%
\bibitem [{\citenamefont {Metzger}\ and\ \citenamefont
  {Karrai}(2004)}]{Metzger2004}%
  \BibitemOpen
  \bibfield  {author} {\bibinfo {author} {\bibfnamefont {C.~H.}\ \bibnamefont
  {Metzger}}\ and\ \bibinfo {author} {\bibfnamefont {K.}~\bibnamefont
  {Karrai}},\ }\href@noop {} {\bibfield  {journal} {\bibinfo  {journal}
  {Nature}\ }\textbf {\bibinfo {volume} {432}},\ \bibinfo {pages} {1002}
  (\bibinfo {year} {2004})}\BibitemShut {NoStop}%
\bibitem [{\citenamefont {Massel}\ \emph {et~al.}(2011)\citenamefont {Massel},
  \citenamefont {Heikkil{\"a}}, \citenamefont {Pirkkalainen}, \citenamefont
  {Cho}, \citenamefont {Saloniemi}, \citenamefont {Hakonen},\ and\
  \citenamefont {Sillanp{\"a}{\"a}}}]{Massel2011}%
  \BibitemOpen
  \bibfield  {author} {\bibinfo {author} {\bibfnamefont {F.}~\bibnamefont
  {Massel}}, \bibinfo {author} {\bibfnamefont {T.~T.}\ \bibnamefont
  {Heikkil{\"a}}}, \bibinfo {author} {\bibfnamefont {J.-M.}\ \bibnamefont
  {Pirkkalainen}}, \bibinfo {author} {\bibfnamefont {S.~U.}\ \bibnamefont
  {Cho}}, \bibinfo {author} {\bibfnamefont {H.}~\bibnamefont {Saloniemi}},
  \bibinfo {author} {\bibfnamefont {P.~J.}\ \bibnamefont {Hakonen}},\ and\
  \bibinfo {author} {\bibfnamefont {M.~A.}\ \bibnamefont {Sillanp{\"a}{\"a}}},\
  }\href@noop {} {\bibfield  {journal} {\bibinfo  {journal} {Nature}\ }\textbf
  {\bibinfo {volume} {480}},\ \bibinfo {pages} {351} (\bibinfo {year}
  {2011})}\BibitemShut {NoStop}%
\bibitem [{\citenamefont {Clerk}\ \emph {et~al.}(2010)\citenamefont {Clerk},
  \citenamefont {Devoret}, \citenamefont {Girvin}, \citenamefont {Marquardt},\
  and\ \citenamefont {Schoelkopf}}]{Clerk2_2010}%
  \BibitemOpen
  \bibfield  {author} {\bibinfo {author} {\bibfnamefont {A.~A.}\ \bibnamefont
  {Clerk}}, \bibinfo {author} {\bibfnamefont {M.~H.}\ \bibnamefont {Devoret}},
  \bibinfo {author} {\bibfnamefont {S.~M.}\ \bibnamefont {Girvin}}, \bibinfo
  {author} {\bibfnamefont {F.}~\bibnamefont {Marquardt}},\ and\ \bibinfo
  {author} {\bibfnamefont {R.~J.}\ \bibnamefont {Schoelkopf}},\ }\href@noop {}
  {\bibfield  {journal} {\bibinfo  {journal} {Rev. Mod. Phys.}\ }\textbf
  {\bibinfo {volume} {82}},\ \bibinfo {pages} {1155} (\bibinfo {year}
  {2010})}\BibitemShut {NoStop}%
\bibitem [{\citenamefont {Fleischhauer}\ \emph {et~al.}(2005)\citenamefont
  {Fleischhauer}, \citenamefont {Imamoglu},\ and\ \citenamefont
  {Marangos}}]{Fleischhauer2005}%
  \BibitemOpen
  \bibfield  {author} {\bibinfo {author} {\bibfnamefont {M.}~\bibnamefont
  {Fleischhauer}}, \bibinfo {author} {\bibfnamefont {A.}~\bibnamefont
  {Imamoglu}},\ and\ \bibinfo {author} {\bibfnamefont {J.~P.}\ \bibnamefont
  {Marangos}},\ }\href@noop {} {\bibfield  {journal} {\bibinfo  {journal} {Rev.
  Mod. Phys.}\ }\textbf {\bibinfo {volume} {77}},\ \bibinfo {pages} {633}
  (\bibinfo {year} {2005})}\BibitemShut {NoStop}%
\bibitem [{\citenamefont {Safavi-Naeini}\ \emph {et~al.}(2011)\citenamefont
  {Safavi-Naeini}, \citenamefont {Alegre}, \citenamefont {Chan}, \citenamefont
  {Eichenfield}, \citenamefont {Winger}, \citenamefont {Lin}, \citenamefont
  {Hill}, \citenamefont {Chang},\ and\ \citenamefont
  {Painter}}]{Safavi-Naeini2011}%
  \BibitemOpen
  \bibfield  {author} {\bibinfo {author} {\bibfnamefont {A.~H.}\ \bibnamefont
  {Safavi-Naeini}}, \bibinfo {author} {\bibfnamefont {T.~P.~M.}\ \bibnamefont
  {Alegre}}, \bibinfo {author} {\bibfnamefont {J.}~\bibnamefont {Chan}},
  \bibinfo {author} {\bibfnamefont {M.}~\bibnamefont {Eichenfield}}, \bibinfo
  {author} {\bibfnamefont {M.}~\bibnamefont {Winger}}, \bibinfo {author}
  {\bibfnamefont {Q.}~\bibnamefont {Lin}}, \bibinfo {author} {\bibfnamefont
  {J.~T.}\ \bibnamefont {Hill}}, \bibinfo {author} {\bibfnamefont {D.~E.}\
  \bibnamefont {Chang}},\ and\ \bibinfo {author} {\bibfnamefont
  {O.}~\bibnamefont {Painter}},\ }\href@noop {} {\bibfield  {journal} {\bibinfo
   {journal} {Nature}\ }\textbf {\bibinfo {volume} {472}},\ \bibinfo {pages}
  {69} (\bibinfo {year} {2011})}\BibitemShut {NoStop}%
\bibitem [{\citenamefont {Rashid}\ \emph {et~al.}(2016)\citenamefont {Rashid},
  \citenamefont {Tufarelli}, \citenamefont {Bateman}, \citenamefont {Vovrosh},
  \citenamefont {Hempston}, \citenamefont {Kim},\ and\ \citenamefont
  {Ulbricht}}]{Hendrik_2016}%
  \BibitemOpen
  \bibfield  {author} {\bibinfo {author} {\bibfnamefont {M.}~\bibnamefont
  {Rashid}}, \bibinfo {author} {\bibfnamefont {T.}~\bibnamefont {Tufarelli}},
  \bibinfo {author} {\bibfnamefont {J.}~\bibnamefont {Bateman}}, \bibinfo
  {author} {\bibfnamefont {J.}~\bibnamefont {Vovrosh}}, \bibinfo {author}
  {\bibfnamefont {D.}~\bibnamefont {Hempston}}, \bibinfo {author}
  {\bibfnamefont {M.~S.}\ \bibnamefont {Kim}},\ and\ \bibinfo {author}
  {\bibfnamefont {H.}~\bibnamefont {Ulbricht}},\ }\href@noop {} {\bibfield
  {journal} {\bibinfo  {journal} {Phys. Rev. Lett.}\ }\textbf {\bibinfo
  {volume} {117}},\ \bibinfo {pages} {273601} (\bibinfo {year}
  {2016})}\BibitemShut {NoStop}%
\bibitem [{\citenamefont {Tan}\ \emph {et~al.}(2013)\citenamefont {Tan},
  \citenamefont {Li},\ and\ \citenamefont {Meystre}}]{Myestre2013}%
  \BibitemOpen
  \bibfield  {author} {\bibinfo {author} {\bibfnamefont {H.}~\bibnamefont
  {Tan}}, \bibinfo {author} {\bibfnamefont {G.}~\bibnamefont {Li}},\ and\
  \bibinfo {author} {\bibfnamefont {P.}~\bibnamefont {Meystre}},\ }\href@noop
  {} {\bibfield  {journal} {\bibinfo  {journal} {Phys. Rev. A}\ }\textbf
  {\bibinfo {volume} {87}},\ \bibinfo {pages} {033829} (\bibinfo {year}
  {2013})}\BibitemShut {NoStop}%
\bibitem [{\citenamefont {Gr{\"o}blacher}\ \emph {et~al.}(2009)\citenamefont
  {Gr{\"o}blacher}, \citenamefont {Hammerer}, \citenamefont {Vanner},\ and\
  \citenamefont {Aspelmeyer}}]{Groblacher2009}%
  \BibitemOpen
  \bibfield  {author} {\bibinfo {author} {\bibfnamefont {S.}~\bibnamefont
  {Gr{\"o}blacher}}, \bibinfo {author} {\bibfnamefont {K.}~\bibnamefont
  {Hammerer}}, \bibinfo {author} {\bibfnamefont {M.~R.}\ \bibnamefont
  {Vanner}},\ and\ \bibinfo {author} {\bibfnamefont {M.}~\bibnamefont
  {Aspelmeyer}},\ }\href@noop {} {\bibfield  {journal} {\bibinfo  {journal}
  {Nature}\ }\textbf {\bibinfo {volume} {460}},\ \bibinfo {pages} {724}
  (\bibinfo {year} {2009})}\BibitemShut {NoStop}%
\bibitem [{\citenamefont {Roelli}\ \emph {et~al.}(2016)\citenamefont {Roelli},
  \citenamefont {Galland}, \citenamefont {Piro},\ and\ \citenamefont
  {Kippenberg}}]{Roelli2016}%
  \BibitemOpen
  \bibfield  {author} {\bibinfo {author} {\bibfnamefont {P.}~\bibnamefont
  {Roelli}}, \bibinfo {author} {\bibfnamefont {C.}~\bibnamefont {Galland}},
  \bibinfo {author} {\bibfnamefont {N.}~\bibnamefont {Piro}},\ and\ \bibinfo
  {author} {\bibfnamefont {T.~J.}\ \bibnamefont {Kippenberg}},\ }\href@noop {}
  {\bibfield  {journal} {\bibinfo  {journal} {Nature Nanotechnology}\ }\textbf
  {\bibinfo {volume} {11}},\ \bibinfo {pages} {164} (\bibinfo {year}
  {2016})}\BibitemShut {NoStop}%
\bibitem [{\citenamefont {Schmidt}\ \emph {et~al.}(2017)\citenamefont
  {Schmidt}, \citenamefont {Esteban}, \citenamefont {Benz}, \citenamefont
  {Baumberg},\ and\ \citenamefont {Aizpurua}}]{Baumberg_2017}%
  \BibitemOpen
  \bibfield  {author} {\bibinfo {author} {\bibfnamefont {M.~K.}\ \bibnamefont
  {Schmidt}}, \bibinfo {author} {\bibfnamefont {R.}~\bibnamefont {Esteban}},
  \bibinfo {author} {\bibfnamefont {F.}~\bibnamefont {Benz}}, \bibinfo {author}
  {\bibfnamefont {J.~J.}\ \bibnamefont {Baumberg}},\ and\ \bibinfo {author}
  {\bibfnamefont {J.}~\bibnamefont {Aizpurua}},\ }\href@noop {} {\bibfield
  {journal} {\bibinfo  {journal} {Faraday Discuss.}\ }\textbf {\bibinfo
  {volume} {205}},\ \bibinfo {pages} {31} (\bibinfo {year} {2017})}\BibitemShut
  {NoStop}%
\bibitem [{\citenamefont {Zhang}\ \emph {et~al.}(2020)\citenamefont {Zhang},
  \citenamefont {Aizpurua},\ and\ \citenamefont {Esteban}}]{Esteban1_2020}%
  \BibitemOpen
  \bibfield  {author} {\bibinfo {author} {\bibfnamefont {Y.}~\bibnamefont
  {Zhang}}, \bibinfo {author} {\bibfnamefont {J.}~\bibnamefont {Aizpurua}},\
  and\ \bibinfo {author} {\bibfnamefont {R.}~\bibnamefont {Esteban}},\
  }\href@noop {} {\bibfield  {journal} {\bibinfo  {journal} {ACS Photonics}\
  }\textbf {\bibinfo {volume} {7}},\ \bibinfo {pages} {1676} (\bibinfo {year}
  {2020})}\BibitemShut {NoStop}%
\bibitem [{\citenamefont {Neuman}\ \emph {et~al.}(2019)\citenamefont {Neuman},
  \citenamefont {Esteban}, \citenamefont {Giedke}, \citenamefont {Schmidt},\
  and\ \citenamefont {Aizpurua}}]{Javier1_2019}%
  \BibitemOpen
  \bibfield  {author} {\bibinfo {author} {\bibfnamefont {T.}~\bibnamefont
  {Neuman}}, \bibinfo {author} {\bibfnamefont {R.}~\bibnamefont {Esteban}},
  \bibinfo {author} {\bibfnamefont {G.}~\bibnamefont {Giedke}}, \bibinfo
  {author} {\bibfnamefont {M.~K.}\ \bibnamefont {Schmidt}},\ and\ \bibinfo
  {author} {\bibfnamefont {J.}~\bibnamefont {Aizpurua}},\ }\href@noop {}
  {\bibfield  {journal} {\bibinfo  {journal} {Phys. Rev. A}\ }\textbf {\bibinfo
  {volume} {100}},\ \bibinfo {pages} {043422} (\bibinfo {year}
  {2019})}\BibitemShut {NoStop}%
\bibitem [{\citenamefont {Schmidt}\ \emph {et~al.}(2016)\citenamefont
  {Schmidt}, \citenamefont {Esteban}, \citenamefont {González-Tudela},
  \citenamefont {Giedke},\ and\ \citenamefont {Aizpurua}}]{Javier2_2016}%
  \BibitemOpen
  \bibfield  {author} {\bibinfo {author} {\bibfnamefont {M.~K.}\ \bibnamefont
  {Schmidt}}, \bibinfo {author} {\bibfnamefont {R.}~\bibnamefont {Esteban}},
  \bibinfo {author} {\bibfnamefont {A.}~\bibnamefont {González-Tudela}},
  \bibinfo {author} {\bibfnamefont {G.}~\bibnamefont {Giedke}},\ and\ \bibinfo
  {author} {\bibfnamefont {J.}~\bibnamefont {Aizpurua}},\ }\href@noop {}
  {\bibfield  {journal} {\bibinfo  {journal} {ACS Nano}\ }\textbf {\bibinfo
  {volume} {10}},\ \bibinfo {pages} {6291} (\bibinfo {year}
  {2016})}\BibitemShut {NoStop}%
\bibitem [{\citenamefont {Esteban}\ \emph {et~al.}(2022)\citenamefont
  {Esteban}, \citenamefont {Baumberg},\ and\ \citenamefont
  {Aizpurua}}]{Esteban2_2022}%
  \BibitemOpen
  \bibfield  {author} {\bibinfo {author} {\bibfnamefont {R.}~\bibnamefont
  {Esteban}}, \bibinfo {author} {\bibfnamefont {J.~J.}\ \bibnamefont
  {Baumberg}},\ and\ \bibinfo {author} {\bibfnamefont {J.}~\bibnamefont
  {Aizpurua}},\ }\href@noop {} {\bibfield  {journal} {\bibinfo  {journal}
  {Accounts of Chemical Research}\ }\textbf {\bibinfo {volume} {55}},\ \bibinfo
  {pages} {1889} (\bibinfo {year} {2022})}\BibitemShut {NoStop}%
\bibitem [{\citenamefont {Tinoco}\ \emph {et~al.}(2013)\citenamefont {Tinoco},
  \citenamefont {Sauer}, \citenamefont {Wang}, \citenamefont {Puglisi},
  \citenamefont {Harbison},\ and\ \citenamefont
  {Rovnyak}}]{tinoco2013physical}%
  \BibitemOpen
  \bibfield  {author} {\bibinfo {author} {\bibfnamefont {I.}~\bibnamefont
  {Tinoco}}, \bibinfo {author} {\bibfnamefont {K.}~\bibnamefont {Sauer}},
  \bibinfo {author} {\bibfnamefont {J.}~\bibnamefont {Wang}}, \bibinfo {author}
  {\bibfnamefont {J.}~\bibnamefont {Puglisi}}, \bibinfo {author} {\bibfnamefont
  {G.}~\bibnamefont {Harbison}},\ and\ \bibinfo {author} {\bibfnamefont
  {D.}~\bibnamefont {Rovnyak}},\ }\href@noop {} {\emph {\bibinfo {title}
  {Physical Chemistry: Principles and Applications in Biological Sciences}}}\
  (\bibinfo  {publisher} {Pearson Education},\ \bibinfo {year}
  {2013})\BibitemShut {NoStop}%
\bibitem [{\citenamefont {Breslow}(2009)}]{breslow2009biomimetic}%
  \BibitemOpen
  \bibfield  {author} {\bibinfo {author} {\bibfnamefont {R.}~\bibnamefont
  {Breslow}},\ }\href@noop {} {\bibfield  {journal} {\bibinfo  {journal}
  {Journal of Biological Chemistry}\ }\textbf {\bibinfo {volume} {284}},\
  \bibinfo {pages} {1337} (\bibinfo {year} {2009})}\BibitemShut {NoStop}%
\bibitem [{\citenamefont {Hong}\ \emph {et~al.}(2010)\citenamefont {Hong},
  \citenamefont {Velegol}, \citenamefont {Chaturvedi},\ and\ \citenamefont
  {Sen}}]{hong2010biomimetic}%
  \BibitemOpen
  \bibfield  {author} {\bibinfo {author} {\bibfnamefont {Y.}~\bibnamefont
  {Hong}}, \bibinfo {author} {\bibfnamefont {D.}~\bibnamefont {Velegol}},
  \bibinfo {author} {\bibfnamefont {N.}~\bibnamefont {Chaturvedi}},\ and\
  \bibinfo {author} {\bibfnamefont {A.}~\bibnamefont {Sen}},\ }\href@noop {}
  {\bibfield  {journal} {\bibinfo  {journal} {Physical Chemistry Chemical
  Physics}\ }\textbf {\bibinfo {volume} {12}},\ \bibinfo {pages} {1423}
  (\bibinfo {year} {2010})}\BibitemShut {NoStop}%
\bibitem [{\citenamefont {Nitzan}(2013)}]{nitzan2013chemical}%
  \BibitemOpen
  \bibfield  {author} {\bibinfo {author} {\bibfnamefont {A.}~\bibnamefont
  {Nitzan}},\ }\bibinfo {title} {Chemical dynamics in condensed phases:
  Relaxation, transfer, and reactions in condensed molecular systems}\
  (\bibinfo  {publisher} {OUP Oxford},\ \bibinfo {year} {2013})\ Chap.~\bibinfo
  {chapter} {10}\BibitemShut {NoStop}%
\bibitem [{\citenamefont {Steck}(2007)}]{steck2007quantum}%
  \BibitemOpen
  \bibfield  {author} {\bibinfo {author} {\bibfnamefont {D.}~\bibnamefont
  {Steck}},\ }\bibinfo {title} {Quantum and atom optics}\ (\bibinfo {year}
  {2007})\ Chap.~\bibinfo {chapter} {12}, pp.\ \bibinfo {pages}
  {509--518}\BibitemShut {NoStop}%
\bibitem [{\citenamefont {Mandal}\ \emph {et~al.}(2020)\citenamefont {Mandal},
  \citenamefont {Krauss},\ and\ \citenamefont {Huo}}]{mandal2020polariton}%
  \BibitemOpen
  \bibfield  {author} {\bibinfo {author} {\bibfnamefont {A.}~\bibnamefont
  {Mandal}}, \bibinfo {author} {\bibfnamefont {T.~D.}\ \bibnamefont {Krauss}},\
  and\ \bibinfo {author} {\bibfnamefont {P.}~\bibnamefont {Huo}},\ }\href@noop
  {} {\bibfield  {journal} {\bibinfo  {journal} {The Journal of Physical
  Chemistry B}\ }\textbf {\bibinfo {volume} {124}},\ \bibinfo {pages} {6321}
  (\bibinfo {year} {2020})}\BibitemShut {NoStop}%
\bibitem [{\citenamefont {F.~Ribeiro}\ \emph {et~al.}(2018)\citenamefont
  {F.~Ribeiro}, \citenamefont {Dunkelberger}, \citenamefont {Xiang},
  \citenamefont {Xiong}, \citenamefont {Simpkins}, \citenamefont {Owrutsky},\
  and\ \citenamefont {Yuen-Zhou}}]{F.Ribeiro2018}%
  \BibitemOpen
  \bibfield  {author} {\bibinfo {author} {\bibfnamefont {R.}~\bibnamefont
  {F.~Ribeiro}}, \bibinfo {author} {\bibfnamefont {A.~D.}\ \bibnamefont
  {Dunkelberger}}, \bibinfo {author} {\bibfnamefont {B.}~\bibnamefont {Xiang}},
  \bibinfo {author} {\bibfnamefont {W.}~\bibnamefont {Xiong}}, \bibinfo
  {author} {\bibfnamefont {B.~S.}\ \bibnamefont {Simpkins}}, \bibinfo {author}
  {\bibfnamefont {J.~C.}\ \bibnamefont {Owrutsky}},\ and\ \bibinfo {author}
  {\bibfnamefont {J.}~\bibnamefont {Yuen-Zhou}},\ }\href@noop {} {\bibfield
  {journal} {\bibinfo  {journal} {The Journal of Physical Chemistry Letters}\
  }\textbf {\bibinfo {volume} {9}},\ \bibinfo {pages} {3766} (\bibinfo {year}
  {2018})}\BibitemShut {NoStop}%
\bibitem [{\citenamefont {Johansson}\ \emph
  {et~al.}(2012{\natexlab{a}})\citenamefont {Johansson}, \citenamefont
  {Nation},\ and\ \citenamefont {Nori}}]{qutip1}%
  \BibitemOpen
  \bibfield  {author} {\bibinfo {author} {\bibfnamefont {J.}~\bibnamefont
  {Johansson}}, \bibinfo {author} {\bibfnamefont {P.}~\bibnamefont {Nation}},\
  and\ \bibinfo {author} {\bibfnamefont {F.}~\bibnamefont {Nori}},\ }\href@noop
  {} {\bibfield  {journal} {\bibinfo  {journal} {Computer Physics
  Communications}\ }\textbf {\bibinfo {volume} {183}},\ \bibinfo {pages} {1760}
  (\bibinfo {year} {2012}{\natexlab{a}})}\BibitemShut {NoStop}%
\bibitem [{\citenamefont {Johansson}\ \emph
  {et~al.}(2012{\natexlab{b}})\citenamefont {Johansson}, \citenamefont
  {Nation},\ and\ \citenamefont {Nori}}]{qutip2}%
  \BibitemOpen
  \bibfield  {author} {\bibinfo {author} {\bibfnamefont {J.}~\bibnamefont
  {Johansson}}, \bibinfo {author} {\bibfnamefont {P.}~\bibnamefont {Nation}},\
  and\ \bibinfo {author} {\bibfnamefont {F.}~\bibnamefont {Nori}},\ }\href@noop
  {} {\bibfield  {journal} {\bibinfo  {journal} {Computer Physics
  Communications}\ }\textbf {\bibinfo {volume} {183}},\ \bibinfo {pages} {1760}
  (\bibinfo {year} {2012}{\natexlab{b}})}\BibitemShut {NoStop}%
\bibitem [{\citenamefont {Werner}\ and\ \citenamefont
  {Hashimoto}(2011)}]{Werner2011}%
  \BibitemOpen
  \bibfield  {author} {\bibinfo {author} {\bibfnamefont {D.}~\bibnamefont
  {Werner}}\ and\ \bibinfo {author} {\bibfnamefont {S.}~\bibnamefont
  {Hashimoto}},\ }\href@noop {} {\bibfield  {journal} {\bibinfo  {journal} {The
  Journal of Physical Chemistry C}\ }\textbf {\bibinfo {volume} {115}},\
  \bibinfo {pages} {5063} (\bibinfo {year} {2011})}\BibitemShut {NoStop}%
\bibitem [{\citenamefont {Johansson}\ \emph {et~al.}(2005)\citenamefont
  {Johansson}, \citenamefont {Xu},\ and\ \citenamefont {K\"all}}]{spectra_3}%
  \BibitemOpen
  \bibfield  {author} {\bibinfo {author} {\bibfnamefont {P.}~\bibnamefont
  {Johansson}}, \bibinfo {author} {\bibfnamefont {H.}~\bibnamefont {Xu}},\ and\
  \bibinfo {author} {\bibfnamefont {M.}~\bibnamefont {K\"all}},\ }\href@noop {}
  {\bibfield  {journal} {\bibinfo  {journal} {Phys. Rev. B}\ }\textbf {\bibinfo
  {volume} {72}},\ \bibinfo {pages} {035427} (\bibinfo {year}
  {2005})}\BibitemShut {NoStop}%
\bibitem [{\citenamefont {Dezfouli}\ \emph {et~al.}(2019)\citenamefont
  {Dezfouli}, \citenamefont {Gordon},\ and\ \citenamefont
  {Hughes}}]{Dezfouli2019}%
  \BibitemOpen
  \bibfield  {author} {\bibinfo {author} {\bibfnamefont {M.~K.}\ \bibnamefont
  {Dezfouli}}, \bibinfo {author} {\bibfnamefont {R.}~\bibnamefont {Gordon}},\
  and\ \bibinfo {author} {\bibfnamefont {S.}~\bibnamefont {Hughes}},\
  }\href@noop {} {\bibfield  {journal} {\bibinfo  {journal} {ACS Photonics}\
  }\textbf {\bibinfo {volume} {6}},\ \bibinfo {pages} {1400} (\bibinfo {year}
  {2019})}\BibitemShut {NoStop}%
\bibitem [{\citenamefont {Dunkelberger}\ \emph
  {et~al.}(2022{\natexlab{b}})\citenamefont {Dunkelberger}, \citenamefont
  {Simpkins}, \citenamefont {Vurgaftman},\ and\ \citenamefont
  {Owrutsky}}]{dunkelberger2022vibration}%
  \BibitemOpen
  \bibfield  {author} {\bibinfo {author} {\bibfnamefont {A.~D.}\ \bibnamefont
  {Dunkelberger}}, \bibinfo {author} {\bibfnamefont {B.~S.}\ \bibnamefont
  {Simpkins}}, \bibinfo {author} {\bibfnamefont {I.}~\bibnamefont
  {Vurgaftman}},\ and\ \bibinfo {author} {\bibfnamefont {J.~C.}\ \bibnamefont
  {Owrutsky}},\ }\href@noop {} {\bibfield  {journal} {\bibinfo  {journal}
  {Annual Review of Physical Chemistry}\ }\textbf {\bibinfo {volume} {73}},\
  \bibinfo {pages} {429} (\bibinfo {year} {2022}{\natexlab{b}})}\BibitemShut
  {NoStop}%
\bibitem [{\citenamefont {Marcus}(1964)}]{marcus1_1964}%
  \BibitemOpen
  \bibfield  {author} {\bibinfo {author} {\bibfnamefont {R.~A.}\ \bibnamefont
  {Marcus}},\ }\href@noop {} {\bibfield  {journal} {\bibinfo  {journal} {Annual
  Review of Physical Chemistry}\ }\textbf {\bibinfo {volume} {15}},\ \bibinfo
  {pages} {155} (\bibinfo {year} {1964})}\BibitemShut {NoStop}%
\bibitem [{\citenamefont {Levich}(1966)}]{levich1966}%
  \BibitemOpen
  \bibfield  {author} {\bibinfo {author} {\bibfnamefont {V.}~\bibnamefont
  {Levich}},\ }\href@noop {} {\bibfield  {journal} {\bibinfo  {journal} {Adv.
  Electrochem. Electrochem. Eng}\ }\textbf {\bibinfo {volume} {4}},\ \bibinfo
  {pages} {249} (\bibinfo {year} {1966})}\BibitemShut {NoStop}%
\bibitem [{\citenamefont {Jortner}(1976)}]{jortner_1976}%
  \BibitemOpen
  \bibfield  {author} {\bibinfo {author} {\bibfnamefont {J.}~\bibnamefont
  {Jortner}},\ }\href@noop {} {\bibfield  {journal} {\bibinfo  {journal} {The
  Journal of Chemical Physics}\ }\textbf {\bibinfo {volume} {64}},\ \bibinfo
  {pages} {4860} (\bibinfo {year} {1976})}\BibitemShut {NoStop}%
\bibitem [{\citenamefont {Abendroth}\ \emph {et~al.}(2015)\citenamefont
  {Abendroth}, \citenamefont {Bushuyev}, \citenamefont {Weiss},\ and\
  \citenamefont {Barrett}}]{pcet1_2015}%
  \BibitemOpen
  \bibfield  {author} {\bibinfo {author} {\bibfnamefont {J.~M.}\ \bibnamefont
  {Abendroth}}, \bibinfo {author} {\bibfnamefont {O.~S.}\ \bibnamefont
  {Bushuyev}}, \bibinfo {author} {\bibfnamefont {P.~S.}\ \bibnamefont
  {Weiss}},\ and\ \bibinfo {author} {\bibfnamefont {C.~J.}\ \bibnamefont
  {Barrett}},\ }\href@noop {} {\bibfield  {journal} {\bibinfo  {journal} {ACS
  nano}\ }\textbf {\bibinfo {volume} {9}},\ \bibinfo {pages} {7746} (\bibinfo
  {year} {2015})}\BibitemShut {NoStop}%
\bibitem [{\citenamefont {Ahlberg}\ \emph {et~al.}(1981)\citenamefont
  {Ahlberg}, \citenamefont {Hammerich},\ and\ \citenamefont
  {Parker}}]{pcet2_1981}%
  \BibitemOpen
  \bibfield  {author} {\bibinfo {author} {\bibfnamefont {E.}~\bibnamefont
  {Ahlberg}}, \bibinfo {author} {\bibfnamefont {O.}~\bibnamefont {Hammerich}},\
  and\ \bibinfo {author} {\bibfnamefont {V.~D.}\ \bibnamefont {Parker}},\
  }\href@noop {} {\bibfield  {journal} {\bibinfo  {journal} {Journal of the
  American Chemical Society}\ }\textbf {\bibinfo {volume} {103}},\ \bibinfo
  {pages} {844} (\bibinfo {year} {1981})}\BibitemShut {NoStop}%
\bibitem [{\citenamefont {Olsen}\ and\ \citenamefont
  {Evans}(1981)}]{pcet3_1981}%
  \BibitemOpen
  \bibfield  {author} {\bibinfo {author} {\bibfnamefont {B.~A.}\ \bibnamefont
  {Olsen}}\ and\ \bibinfo {author} {\bibfnamefont {D.~H.}\ \bibnamefont
  {Evans}},\ }\href@noop {} {\bibfield  {journal} {\bibinfo  {journal} {Journal
  of the American Chemical Society}\ }\textbf {\bibinfo {volume} {103}},\
  \bibinfo {pages} {839} (\bibinfo {year} {1981})}\BibitemShut {NoStop}%
\bibitem [{\citenamefont {Campos-Gonzalez-Angulo}\ \emph
  {et~al.}(2019)\citenamefont {Campos-Gonzalez-Angulo}, \citenamefont
  {Ribeiro},\ and\ \citenamefont {Yuen-Zhou}}]{jorge2019}%
  \BibitemOpen
  \bibfield  {author} {\bibinfo {author} {\bibfnamefont {J.~A.}\ \bibnamefont
  {Campos-Gonzalez-Angulo}}, \bibinfo {author} {\bibfnamefont {R.~F.}\
  \bibnamefont {Ribeiro}},\ and\ \bibinfo {author} {\bibfnamefont
  {J.}~\bibnamefont {Yuen-Zhou}},\ }\href@noop {} {\bibfield  {journal}
  {\bibinfo  {journal} {Nature Communications}\ }\textbf {\bibinfo {volume}
  {10}},\ \bibinfo {pages} {4685} (\bibinfo {year} {2019})}\BibitemShut
  {NoStop}%
\bibitem [{\citenamefont {Colbert}\ and\ \citenamefont
  {Miller}(1992)}]{colbert_1992}%
  \BibitemOpen
  \bibfield  {author} {\bibinfo {author} {\bibfnamefont {D.~T.}\ \bibnamefont
  {Colbert}}\ and\ \bibinfo {author} {\bibfnamefont {W.~H.}\ \bibnamefont
  {Miller}},\ }\href@noop {} {\bibfield  {journal} {\bibinfo  {journal} {The
  Journal of Chemical Physics}\ }\textbf {\bibinfo {volume} {96}},\ \bibinfo
  {pages} {1982} (\bibinfo {year} {1992})}\BibitemShut {NoStop}%
\bibitem [{\citenamefont {Hopfield}(1969)}]{hopfield1969}%
  \BibitemOpen
  \bibfield  {author} {\bibinfo {author} {\bibfnamefont {J.~J.}\ \bibnamefont
  {Hopfield}},\ }\href@noop {} {\bibfield  {journal} {\bibinfo  {journal}
  {Phys. Rev.}\ }\textbf {\bibinfo {volume} {182}},\ \bibinfo {pages} {945}
  (\bibinfo {year} {1969})}\BibitemShut {NoStop}%
\bibitem [{\citenamefont {del Pino}\ \emph {et~al.}(2015)\citenamefont {del
  Pino}, \citenamefont {Feist},\ and\ \citenamefont
  {Garcia-Vidal}}]{ohmic_2015}%
  \BibitemOpen
  \bibfield  {author} {\bibinfo {author} {\bibfnamefont {J.}~\bibnamefont {del
  Pino}}, \bibinfo {author} {\bibfnamefont {J.}~\bibnamefont {Feist}},\ and\
  \bibinfo {author} {\bibfnamefont {F.~J.}\ \bibnamefont {Garcia-Vidal}},\
  }\href@noop {} {\bibfield  {journal} {\bibinfo  {journal} {New Journal of
  Physics}\ }\textbf {\bibinfo {volume} {17}},\ \bibinfo {pages} {053040}
  (\bibinfo {year} {2015})}\BibitemShut {NoStop}%
\bibitem [{\citenamefont {Xiang}\ \emph {et~al.}(2018)\citenamefont {Xiang},
  \citenamefont {Ribeiro}, \citenamefont {Dunkelberger}, \citenamefont {Wang},
  \citenamefont {Li}, \citenamefont {Simpkins}, \citenamefont {Owrutsky},
  \citenamefont {Yuen-Zhou},\ and\ \citenamefont {Xiong}}]{xiang_2018}%
  \BibitemOpen
  \bibfield  {author} {\bibinfo {author} {\bibfnamefont {B.}~\bibnamefont
  {Xiang}}, \bibinfo {author} {\bibfnamefont {R.~F.}\ \bibnamefont {Ribeiro}},
  \bibinfo {author} {\bibfnamefont {A.~D.}\ \bibnamefont {Dunkelberger}},
  \bibinfo {author} {\bibfnamefont {J.}~\bibnamefont {Wang}}, \bibinfo {author}
  {\bibfnamefont {Y.}~\bibnamefont {Li}}, \bibinfo {author} {\bibfnamefont
  {B.~S.}\ \bibnamefont {Simpkins}}, \bibinfo {author} {\bibfnamefont {J.~C.}\
  \bibnamefont {Owrutsky}}, \bibinfo {author} {\bibfnamefont {J.}~\bibnamefont
  {Yuen-Zhou}},\ and\ \bibinfo {author} {\bibfnamefont {W.}~\bibnamefont
  {Xiong}},\ }\href@noop {} {\bibfield  {journal} {\bibinfo  {journal}
  {Proceedings of the National Academy of Sciences}\ }\textbf {\bibinfo
  {volume} {115}},\ \bibinfo {pages} {4845} (\bibinfo {year}
  {2018})}\BibitemShut {NoStop}%
\bibitem [{\citenamefont {Xiang}\ \emph {et~al.}(2020)\citenamefont {Xiang},
  \citenamefont {Ribeiro}, \citenamefont {Du}, \citenamefont {Chen},
  \citenamefont {Yang}, \citenamefont {Wang}, \citenamefont {Yuen-Zhou},\ and\
  \citenamefont {Xiong}}]{xiang2020intermolecular}%
  \BibitemOpen
  \bibfield  {author} {\bibinfo {author} {\bibfnamefont {B.}~\bibnamefont
  {Xiang}}, \bibinfo {author} {\bibfnamefont {R.~F.}\ \bibnamefont {Ribeiro}},
  \bibinfo {author} {\bibfnamefont {M.}~\bibnamefont {Du}}, \bibinfo {author}
  {\bibfnamefont {L.}~\bibnamefont {Chen}}, \bibinfo {author} {\bibfnamefont
  {Z.}~\bibnamefont {Yang}}, \bibinfo {author} {\bibfnamefont {J.}~\bibnamefont
  {Wang}}, \bibinfo {author} {\bibfnamefont {J.}~\bibnamefont {Yuen-Zhou}},\
  and\ \bibinfo {author} {\bibfnamefont {W.}~\bibnamefont {Xiong}},\
  }\href@noop {} {\bibfield  {journal} {\bibinfo  {journal} {Science}\ }\textbf
  {\bibinfo {volume} {368}},\ \bibinfo {pages} {665} (\bibinfo {year}
  {2020})}\BibitemShut {NoStop}%
\bibitem [{\citenamefont {Morris}(2012)}]{morris2012dawn}%
  \BibitemOpen
  \bibfield  {author} {\bibinfo {author} {\bibfnamefont {C.}~\bibnamefont
  {Morris}},\ }\bibinfo {title} {The dawn of innovation: The first american
  industrial revolution}\ (\bibinfo {year} {2012})\ p.~\bibinfo {pages}
  {42}\BibitemShut {NoStop}%
\bibitem [{\citenamefont {Richards}(2016)}]{Richards2016}%
  \BibitemOpen
  \bibfield  {author} {\bibinfo {author} {\bibfnamefont {V.}~\bibnamefont
  {Richards}},\ }\href@noop {} {\bibfield  {journal} {\bibinfo  {journal}
  {Nature Chemistry}\ }\textbf {\bibinfo {volume} {8}},\ \bibinfo {pages}
  {1090} (\bibinfo {year} {2016})}\BibitemShut {NoStop}%
\end{thebibliography}%
\end{document}


\title{Supplementary information: Thermodynamic coupling of reactions via few-molecule vibrational polaritons}

\author{Arghadip Koner}
\affiliation{Department of Chemistry and Biochemistry, University of California
San Diego, La Jolla, California 92093, USA}
\author{Matthew Du}
\affiliation{Department of Chemistry, University of Chicago, 5735 S Ellis Ave, Chicago, Illinois 60637, USA}
\author{Sindhana Pannir-Sivajothi}
\affiliation{Department of Chemistry and Biochemistry, University of California
San Diego, La Jolla, California 92093, USA}
\author{Randall H. Goldsmith}
\affiliation{Department of Chemistry, University of Wisconsin-Madison, Madison, Wisconsin 53706-1322, USA}
\author{Joel Yuen-Zhou}
\email{joelyuen@ucsd.edu}
\affiliation{Department of Chemistry and Biochemistry, University of California
San Diego, La Jolla, California 92093, USA}

\maketitle
\subsection*{Supplementary note 1: Effective Hamiltonian}
The full Hamiltonian of the optomechanical setup with the molecular vibrational mode $(b^\dagger,b)$ at frequency $\Omega_{\text{v}}$ and the bare cavity $(a^\dagger,a)$ at frequency $\omega_{\text{cav}}$, driven by a laser modeled as a quantum harmonic oscillator 
$(l,l^{\dagger})$ with frequency $\omega_{\text{L}}$ is given as

\begin{align*}
H_{\text{full}} & =H_{\text{mol}}+H_{\text{cav}}+H_{\text{laser}}+H_{\text{cm}}+H_{\text{lc}},\\
 & =\hbar\Omega_{\text{v}}b^{\dagger}b+\hbar\omega_{\text{cav}}a^{\dagger}a+\hbar\omega_{\text{L}}l^{\dagger}l+\hbar g_{0}a^{\dagger}a(b^{\dagger}+b)+\hbar J(a^{\dagger}+a)(l^{\dagger}+l),
\end{align*}
With the rotating wave approximation (RWA) in the laser-cavity coupling, we have
\begin{eqnarray}
H^{\text{RWA}}_{\text{full}}=\hbar\Omega_{\text{v}}b^{\dagger}b+\hbar\omega_{\text{cav}}a^{\dagger}a+\hbar\omega_{\text{L}}l^{\dagger}l+\hbar g_{0}a^{\dagger}a(b^{\dagger}+b)+\hbar J(a^{\dagger}l+al^{\dagger}).
\end{eqnarray}
Since the cavity and the laser are linearly coupled, we can first diagonalize this part of the Hamiltonian,
\begin{eqnarray}
\nonumber H_{\text{L}-\text{C}}&=&\hbar\omega_{\text{L}}l^{\dagger}l+\hbar\omega_{\text{cav}}a^{\dagger}a+\hbar J(a^{\dagger}l+al^{\dagger}), \\
&=& \hbar\tilde{\omega}_{\text{L}}\tilde{a}^{\dagger}\tilde{a}+\hbar\tilde{\omega}_{\text{cav}}\tilde{l}^{\dagger}\tilde{l},
\end{eqnarray}
where
\begin{align*}
\tilde{l}=\sin\varphi\cdot a+\cos\varphi\cdot l,\quad\tilde{\omega}_{\text{L}} & =\frac{(\omega_{\text{cav}}+\omega_{\text{L}})-\sqrt{4J^{2}+(\omega_{\text{cav}}-\omega_{\text{L}})^{2}}}{2},\\
\tilde{a}=\cos\varphi\cdot a-\sin\varphi\cdot l,\quad\tilde{\omega}_{\text{cav}} & =\frac{(\omega_{\text{cav}}+\omega_{\text{L}})+\sqrt{4J^{2}+(\omega_{\text{cav}}-\omega_{\text{L}})^{2}}}{2}.
\end{align*}
Here, $\varphi=\frac{1}{2}\tan^{-1}\bigg(\frac{2J}{\omega_{\text{cav}}-\omega_{\text{L}}}\bigg)$ is the mixing angle.
For small $J$, $\tilde{l}$ is a `laser-like' mode and $\tilde{a}$
is a `cavity like' mode. Rewriting $H^{\text{RWA}}_\text{full}$ in the new normal
mode basis

\[
H^{\text{RWA}}_{\text{full}}=\hbar\Omega_{\text{v}}b^{\dagger}b+\hbar\tilde{\omega}_{\text{cav}}\tilde{a}^{\dagger}\tilde{a}+\hbar\tilde{\omega}_{\text{L}}\tilde{l}^{\dagger}\tilde{l}+\hbar g_{0}\big(\cos\varphi\cdot\tilde{a}+\sin\varphi\cdot\tilde{l}\big)^{\dagger}\big(\cos\varphi\cdot\tilde{a}+\sin\varphi\cdot\tilde{l}\big)(b^{\dagger}+b).
\]
For the laser being red detuned (${\omega}_{\text{L}}={\omega}_{\text{cav}}-\Delta$, $\Delta > 0$),
  keeping only the near
resonant terms, we have 
\[
H_{\text{R}}=\hbar\Omega_{\text{v}}b^{\dagger}b+\hbar\tilde{\omega}_{\text{cav}}\tilde{a}^{\dagger}\tilde{a}+\hbar\tilde{\omega}_{\text{L}}\tilde{l}^{\dagger}\tilde{l}+\hbar g_{0}\cos\varphi\sin\varphi\cdot(\tilde{l}{}^{\dagger}\tilde{a}b^{\dagger}+\tilde{l}\tilde{a}^{\dagger}b).
\]
The Heisenberg equation of motion (EOM) for $(\tilde{l}^{\dagger}\tilde{a})$ is
\[
\frac{d}{dt}(\tilde{l}^{\dagger}\tilde{a})=-i(\tilde{\omega}_{\text{cav}}-\tilde{\omega}_{\text{L}})\tilde{l}^{\dagger}\tilde{a}+ig_{0}\cos\varphi\sin\varphi\cdot\big[\tilde{l}\tilde{a}^{\dagger},\tilde{l}^{\dagger}\tilde{a}\big]\cdot b.
\]
Computing the commutator
\begin{align*}
\big[\tilde{l}\tilde{a}^{\dagger},\tilde{l}^{\dagger}\tilde{a}\big] & =\tilde{l}\tilde{a}^{\dagger}\tilde{l}^{\dagger}\tilde{a}-\tilde{l}^{\dagger}\tilde{a}\tilde{l}\tilde{a}^{\dagger}.\\
\text{Using,\ensuremath{\quad}}\tilde{a}\tilde{a}^{\dagger}=\tilde{a}^{\dagger}\tilde{a}+1\\
 & =\tilde{l}\tilde{a}^{\dagger}\tilde{l}^{\dagger}\tilde{a}-\tilde{l}^{\dagger}\tilde{l}(\tilde{a}^{\dagger}\tilde{a}+1)\\
 & =\tilde{a}^{\dagger}\tilde{a}\cdot[\tilde{l},\tilde{l}^{\dagger}]-\tilde{l}^{\dagger}\tilde{l}\\
 & =\tilde{a}^{\dagger}\tilde{a}-\tilde{l}^{\dagger}\tilde{l}.
\end{align*}
Then,
\begin{align}\label{EOM_la}
\nonumber \frac{d}{dt}(\tilde{l}^{\dagger}\tilde{a}) & =-i(\tilde{\omega}_{\text{cav}}-\tilde{\omega}_{\text{L}})\tilde{l}^{\dagger}\tilde{a}+ig_{0}\cos\varphi\sin\varphi\cdot(\tilde{a}^{\dagger}\tilde{a}-\tilde{l}^{\dagger}\tilde{l})\cdot b,\\
 & =-i(\tilde{\omega}_{\text{cav}}-\tilde{\omega}_{\text{L}})\tilde{l}^{\dagger}\tilde{a}-ig_{0}\cos\varphi\sin\varphi\cdot(\tilde{l}^{\dagger}\tilde{l}-\tilde{a}^{\dagger}\tilde{a})\cdot b.
\end{align}
We will now make the mean-field approximation to linearize the equation
of motion. For the three-body operators of the form $c^{\dagger}cb$
(where $c=\tilde{l}$ or $\tilde{a}$), we have
\begin{align*}
c^{\dagger}cb & =(\underbrace{\langle c^{\dagger}c\rangle}_{\text{mean}}+\underbrace{c^{\dagger}c-\langle c^{\dagger}c\rangle}_{\text{Fluctuations}})\cdot b,\\
 & =\langle c^{\dagger}c\rangle\cdot b+(c^{\dagger}c-\langle c^{\dagger}c\rangle)\cdot b,\\
 & \approx\langle c^{\dagger}c\rangle b.
\end{align*}
Here we have neglected the fluctuations in $\langle c^{\dagger}c\rangle$. Choosing, 
\[
\langle\tilde{l}^{\dagger}\tilde{l}\rangle=\tilde{n}_{\text{L}},\langle\tilde{a}^{\dagger}\tilde{a}\rangle=\tilde{n}_{a},
\]
and plugging these back into the equation~\ref{EOM_la}
\begin{align}
\frac{d}{dt}(\tilde{l}^{\dagger}\tilde{a}) & =-i(\tilde{\omega}_{\text{cav}}-\tilde{\omega}_{\text{L}})\tilde{l}^{\dagger}\tilde{a}-ig_{0}\cos\varphi\sin\varphi\cdot(\tilde{n}_{\text{L}}-\tilde{n}_{a})\cdot b.
\end{align}
Now writing down the EOM for $b$,
\begin{eqnarray}
\frac{d}{dt}b=-i\Omega_{\text{v}}b-ig_{0}\cos\varphi\sin\varphi\cdot\big(\tilde{l}^{\dagger}\tilde{a}).
\end{eqnarray}
To write an effective Hamiltonian, we define a composite mode for the laser-cavity subsystem $\mathcal{A}_{\text{ph}}\equiv\frac{\tilde{l}^{\dagger}\tilde{a}}{\sqrt{(\tilde{n}_{\text{L}}-\tilde{n}_{a})}}$. The EOM for operators $\mathcal{A}_{\text{ph}}$ and $b$ are
\begin{align*}
\frac{d}{dt}\mathcal{A}_{\text{ph}} & =-i(\tilde{\omega}_{\text{cav}}-\tilde{\omega}_{\text{L}}) \mathcal{A}_{\text{ph}} -ig_{0}\cos\varphi\sin\varphi\cdot\sqrt{(\tilde{n}_{\text{L}}-\tilde{n}_{a})}\cdot b,\\
\frac{d}{dt}b & =-i\Omega_{\text{v}}b-ig_{0}\cos\varphi\sin\varphi\sqrt{(\tilde{n}_{\text{L}}-\tilde{n}_{a})}\cdot\mathcal{A}_{\text{ph}}.
\end{align*}
For $J\ll\Delta$, we have $\tilde{\omega}_{\text{L}}\approx\tilde{\omega}_{\text{cav}}-\Delta$ and $\tilde{n}_{\text{L}}\approx \langle{l^\dagger l}\rangle \equiv n_{\text{L}}$. Also
\begin{align*}
\cos\varphi\sin\varphi & =\frac{1}{2}\sin2\varphi\\
 & =\frac{1}{2}\sin\bigg(\tan^{-1}\bigg(\frac{2J}{\omega_{\text{cav}}-\omega_{\text{L}}}\bigg)\bigg)\\
 & =\frac{J}{\sqrt{\Delta^{2}+4J^{2}}} \approx \frac{J}{\Delta}. 
\end{align*}
In this limit, the EOMs transform to
\begin{align*}
\frac{d}{dt}\mathcal{A}_{\text{ph}} & =-i\Delta \mathcal{A}_{\text{ph}} -ig_{0}\bigg( \frac{J}{\Delta}\bigg)\cdot\sqrt{({n}_{\text{L}}-\tilde{n}_{a})}\cdot b,\\
\frac{d}{dt}b & =-i\Omega_{\text{v}}b-ig_{0}\bigg( \frac{J}{\Delta}\bigg)\sqrt{({n}_{\text{L}}-\tilde{n}_{a})}\cdot\mathcal{A}_{\text{ph}}.
\end{align*}
Now using the fact that $n_{\text{L}}\gg \tilde{n}_{a}$, we have

\begin{align*}
\frac{d}{dt}\mathcal{A}_{\text{ph}} & =-i\Delta \mathcal{A}_{\text{ph}} -ig_{0}\bigg( \frac{J}{\Delta}\bigg)\cdot\sqrt{n_{\text{L}}}\sqrt{1-\frac{\tilde{n}_{a}}{n_{\text{L}}}}\cdot b,&\\
&\approx -i\Delta \mathcal{A}_{\text{ph}} -ig_{0}\bigg( \frac{J}{\Delta}\bigg)\sqrt{n_{\text{L}}}\cdot b,&\\
\frac{d}{dt}b & =-i\Omega_{\text{v}}b-ig_{0}\bigg( \frac{J}{\Delta}\bigg)\sqrt{{n}_{\text{L}}}\cdot\mathcal{A}_{\text{ph}}.
\end{align*}
These EOMs look like two coupled oscillators with coupling constant
$g_{0}\big( \frac{J}{\Delta}\big)\sqrt{{n}_{\text{L}}}$. Thus, we can write an
effective Hamiltonian for this system as

\begin{align}
H_{\text{eff}} & =\hbar\Delta \mathcal{A}_{\text{ph}}^{\dagger} \mathcal{A}_{\text{ph}}+\hbar\Omega_{\text{v}}b^{\dagger}b+\hbar g_{0}\bigg( \frac{J}{\Delta}\bigg)\sqrt{n_{\text{L}}}\big( \mathcal{A}_{\text{ph}}^{\dagger}b+ \mathcal{A}_{\text{ph}}b^{\dagger}\big)\label{eq:8}.
\end{align}
We note that for $J\ll\Delta$ and ${n}_{\text{L}}\gg\tilde{n}_a$, $\{ \mathcal{A}_{\text{ph}}, \mathcal{A}_{\text{ph}}^{\dagger}\}$
satisfy bosonic commutation relations,

\begin{align*}
[ \mathcal{A}_{\text{ph}}, \mathcal{A}_{\text{ph}}^{\dagger}] & =\frac{\tilde{l}^{\dagger}\tilde{a}\tilde{l}\tilde{a}^{\dagger}-\tilde{l}\tilde{a}^{\dagger}\tilde{l}^{\dagger}\tilde{a}}{(\tilde{n}_{\text{L}}-\tilde{n}_{a})},\\
 & = \frac{\tilde{l}^{\dagger}\tilde{l}-\tilde{a}^{\dagger}\tilde{a}}{{(\tilde{n}_{\text{L}}-\tilde{n}_{a})}},\\
 & \approx \mathbb{I}.
\end{align*}

	\subsection*{Supplementary note 2: Decay rate for the composite boson}
The full Hamiltonian in the RWA with the decay of the cavity and the
vibrational mode as $\kappa$ and $\gamma$ (we are assuming that
the laser mode has no incohorent decay) is given as

\[
H=\hbar\bigg(\Omega_{\text{v}}-i\frac{\gamma}{2}\bigg)b^{\dagger}b+\hbar\bigg(\omega_{\text{cav}}-i\frac{\kappa}{2}\bigg)a^{\dagger}a+\hbar\omega_{\text{L}}l^{\dagger}l+\hbar g_{0}a^{\dagger}a(b^{\dagger}+b)+\hbar J(a^{\dagger}l+al^{\dagger}).
\]
Diagonalizing the cavity-laser subsystem

\[
H_{\text{L}-\text{C}}=\hbar\bigg(\omega_{\text{cav}}-i\frac{\kappa}{2}\bigg)a^{\dagger}a+\hbar\omega_{\text{L}}l^{\dagger}l+\hbar J(a^{\dagger}l+al^{\dagger}),
\]
the normal mode frequencies are
\begin{align*}
\tilde{\omega}_{\text{L}} & =\frac{\omega_{\text{L}}+(\omega_{\text{cav}}-i\kappa/2)-\sqrt{4J^{2}+[(\omega_{\text{cav}}-i\kappa/2) - \omega_{\text{L}}]^{2}}}{2},\\
\tilde{\omega}_{\text{cav}} & =\frac{\omega_{\text{L}}+(\omega_{\text{cav}}-i\kappa/2)+\sqrt{4J^{2}+[(\omega_{\text{cav}}-i\kappa/2) - \omega_{\text{L}}]^{2}}}{2}.
\end{align*}
Now, we need the decay for the composite bosons. Let's consider the
Hamiltonian 

\[
H_{0}=\hbar\tilde{\omega}_{\text{L}}\tilde{l}^{\dagger}\tilde{l}+\hbar\tilde{\omega}_{\text{cav}}\tilde{a}^{\dagger}\tilde{a}.
\]
In the Heisenberg picture

\[
\tilde{l}(t)=\tilde{l}e^{-i\tilde{\omega}_{\text{L}}t},\quad\tilde{a}(t)=\tilde{a}e^{-i\tilde{\omega}_{\text{cav}}t}.
\]
The Heisenberg EOM for $(\tilde{l}^{\dagger}\tilde{a})$ is 

\begin{align*}
\frac{d}{dt}(\tilde{l}^{\dagger}\tilde{a}) & =\frac{i}{\hbar}[H_{0},\tilde{l}^{\dagger}\tilde{a}]\\
 & =-i(\tilde{\omega}_{\text{cav}}-\tilde{\omega}_{\text{L}})\tilde{l}^{\dagger}\tilde{a}.
\end{align*}
Thus

\begin{align*}
\tilde{l}^{\dagger}\tilde{a}(t) & =\tilde{l}^{\dagger}\tilde{a}e^{-i(\tilde{\omega}_{\text{cav}}-\tilde{\omega}_{\text{L}})t},
\end{align*}
where
\begin{align*}
\tilde{\omega}_{\text{cav}}-\tilde{\omega}_{\text{L}}  = & \frac{(\omega_{\text{cav}}-i\kappa/2)+\omega_{\text{L}}+\sqrt{4J^{2}+[(\omega_{\text{cav}}-i\kappa/2) - \omega_{\text{L}}]{}^{2}}}{2}-& \\  &\frac{(\omega_{\text{cav}}-i\kappa/2)+\omega_{\text{L}}-\sqrt{4J^{2}+[(\omega_{\text{cav}}-i\kappa/2) - \omega_{\text{L}}]{}^{2}}}{2},\\
 & =\sqrt{4J^{2}+[(\omega_{\text{cav}}-i\kappa/2) - \omega_{\text{L}}]{}^{2}},&\\
 & \approx(\omega_{\text{cav}}-\omega_{\text{L}})-i\kappa/2,
\end{align*}
for $J\ll(\omega_{\text{cav}}-\omega_{\text{L}})$.\\

Thus, we show that the incoherent decay rate for the composite boson is the same as that of
the cavity decay assuming that the laser mode has no incoherent decay. The full Hamiltonian in the normal mode basis
is then given as

\[
H=\hbar\bigg(\Omega_{\text{v}}-i\frac{\gamma}{2}\bigg)b^{\dagger}b+\hbar\bigg((\omega_{\text{cav}}-\omega_{\text{L}})-i\frac{\kappa}{2}\bigg) \mathcal{A}_{\text{ph}}^{\dagger} \mathcal{A}_{\text{ph}}+\hbar g_{0}\bigg( \frac{J}{\Delta}\bigg)\sqrt{{n}_{\text{L}}}\big( \mathcal{A}_{\text{ph}}^{\dagger}b+ \mathcal{A}_{\text{ph}}b^{\dagger}\big).
\]

\bibliographystyle{unsrtnat}
\bibliography{}